\documentclass[aps,pre,preprint,groupedaddress]{revtex4}
\usepackage[normalem]{ulem}
\usepackage{float}
\usepackage[usenames]{color}
\usepackage{graphicx}
\usepackage{ulem}
\usepackage{amsfonts}
\usepackage{amsmath}
\usepackage{subfigure}
\newcommand{\um}{\ensuremath{\mu \text{m}\,\,}}

\newcommand{\degree}{\ensuremath{^{\circ}}}
\newcommand{\avg}[1]{\langle #1 \rangle}
\newcommand{\pd}[2]{\frac{\partial #1}{\partial #2}}

\begin{document}

\title{Automated three-dimensional single cell phenotyping of spindle dynamics, cell shape, and volume}
% Quantitative Analysis of Cell Nucleus Organization
\author{Kemp Plumb}
\affiliation{Department of Physics and Astronomy, Johns Hopkins University, Baltimore, Maryland 21218, USA}
\author{Sarah Elaz}
\affiliation{Department of Physics, University of Massachusetts, Amherst, Amherst, Massachusetts 01003, USA}
\author{Vincent Pelletier}
\affiliation{Quorum Technologies Inc., Guelph, Ontario, Canada}
\author{Maria L. Kilfoil}
\affiliation{Department of Physics, University of Massachusetts Amherst, Amherst, Massachusetts 01003, USA}

\date{\today}

\begin{abstract}
We present feature finding and tracking algorithms in 3D in living cells, and demonstrate their utility to measure metrics important in cell biological processes. We developed a computational imaging hybrid approach that combines automated three-dimensional tracking of point-like features, with surface reconstruction from which cell (or nuclear) volume, shape, and planes of interest can be extracted. After validation, we applied the technique to real space context-rich dynamics of the mitotic spindle, and cell volume and its relationship to spindle length, in dividing living cells. These methods are additionally useful for automated segregation of pre-anaphase and anaphase spindle populations. We found that genetic deletion of the yeast kinesin-5 mitotic motor cin8 leads to abnormally large mother and daughter cells that were indistinguishable based on size, and that in those cells the spindle length becomes  uncorrelated with cell size. The technique can be used to visualize and quantify tracked feature coordinates relative to cell bounding surface landmarks, such as the plane of cell division. By virtue of enriching the information content of subcellular dynamics, these tools will help advance studies of non-equilibrium processes in living cells.
\end{abstract}

\maketitle
Locating submicrometer entities at their address in a living cell is an important task in capturing subcellular dynamics. 
Current imaging studies of subcellular, or subnuclear, dynamics of localized features of interest in small bounding volumes, for example in the important model organism budding yeast, the eukaryotic nucleus, or ER / Golgi bodies, suffer from important limitations. First, dynamics single features such as a chromosome locus position 
are often described only by position, without contextual information~\cite{Weber-PNAS-2012,CourtyDahan-2006}, subjecting the extracted features to translational, rotational, or vibrational noise, and to lack of knowledge of position in the cell's cycle.  Features that are free to move in three-dimensional (3D) space are often either detected in two dimensions (2D)~\cite{Weber-PNAS-2012}, or in 3D without subpixel resolution in 3D, allowing out-of-plane motion to be ascribed to in-plane motion. In many other studies, contextual information is used, for example relative positions of two spots such as in the spindle pole bodies (SPBs), but without subpixel resolution in 3D~\cite{Straight_Science_97}, so that the uncertainty in the measurement is insufficiently small to enable new and valuable dynamical information at tens of nanometers scale to be uncovered. There are however important exceptions to these limitations~\cite{Gasser_PNAS04}. 
Finally, in many reports cell volume is extracted by measuring an area and extrapolating to volume by assuming the cell has some regular azimuthal symmetry~\cite{Plaza-volume-PRE2014}, which it may not have. 
 In various micropipette aspiration studies to determine mechanical behavior, the cell has been either considered as a fully incompressible material (with true 3D volume unmonitored during deformation)~\cite{Robinson-2013}, or has been assumed compressible without any emphasis on its importance~\cite{Trickey-Guilak-2006}. 
Yet, a number of cell types have already been shown to be compressible~\cite{Zonal-chondocyte-3Dvolume,FE-compressibility,LiChen-FEcompressibility}. Cell compressibility may prove to be more general~\cite{Guilak1995,GuilakRatcliffe1995}. On account of these limitations, changes in cell volume due to deformation and other perturbations may be strongly underappreciated. 
Commercial software solutions exist for extracting volume using isosurface methods on a continuous mass distribution and have been applied to cells~\cite{Zonal-chondocyte-3Dvolume,compressibililty-chondrocyte-2011}.  Isosurface methods are notoriously sensitive to the threshold chosen.  There is a need for methods of reconstructing cell or small subcellular volumes surface that are relatively insensitive to thresholding, and that are robust to challenging conditions of local regions of high curvature. 

%Three-dimensional shapes of chondrocytes were reconstructed using the commercial software (Image-Pro V6.3, Media Cybernetics, Inc., USA).
Clearly, additional, and more flexible and accessible, high-accuracy computational imaging tools for point and bounded surface features are needed. 
%%%%% CONTEXT dependent motion Not only obtain high resolution in 3D, the space in which objects move, but  determine where in the cell cycle specific dynamics are taking place. 
Such methods would not only enable direct volume measurements in combination with feature tracking of localized entities, but would facilitate accurate monitoring of cell volume under nonlinear deformation of cells, and open up new avenues distinguishing 
between compressibility and viscoelastic bulk relaxation/fluid flow in the nonlinear deformation behavior of cells.
 %~\cite{FE-compressibility}.

%We have solved this problem of computing the bounding surface that approximates the original surface in dividing budding yeast cells, 
%using a water-tight method that is robust to local high surface curvature. 
%The method permits simultaneous three-dimensional monitoring during cell during division of the volumes of dividing cells and the orientation of the virtual plane separating them, together with dynamics of any features separately detected and tracked in parallel in the cell. 
Here we describe a hybrid computational imaging technique that overcomes these obstacles and generates high resolution tracking of nanoscale to organelle-sized spots in living cells at identifiable locations within the cell, together with bounding surfaces, from hundreds of cells. The watertight bounding surface is obtained using computational geometry techniques and enables monitoring of the volumes of dividing cells and the orientation of the virtual plane separating them. The features can then be transformed to a new co-ordinate system unique to each cell if desired. These methods allow for accurate dissection of yeast mitosis or chromosome locus motions in their precise cellular context {\em in vivo}. 
%
%Our development of a novel surface hybrid of high-resolution tracking of coordinates of multiple features, and 
%%%quantitative motions or 
%volumes of individual small biological cells or subcellular volumes that can be fluorescently labelled, permits 
%measurement of nanometer scale dynamics, relative dynamics, and dynamics relative to a reference point on the surface, of nanometer scale structures.  
Besides automated high-resolution, high-throughput image and data analysis, our technique also makes possible the extraction of new phenotypic metrics by which cell-to-cell heterogeneity across populations can by quantified~\cite{Manning-contour-based-2014}.  Much cellular motion is stochastic and must be assessed statistically from large populations. 
The ability to calculate volumes and shapes in situations of low sampling of surface points is likely to become more important as more sophisticated models are developed for  how the interior of the cell and nucleus are organized. 

Here, we use these methods to demonstrate nanometer-scale measurements of directed motions or fluctuations of the eukaryotic mitotic spindle, using relative motion to unambiguously identify cells in different stages of cell division; additionally, to demonstrate measurements of cell volume of individual small biological cells.
We show that the resolution of the point feature finding in our optical system, for which we perform simulations of model spindle poles under similar image detection conditions to what is expected in the living cell, is $~10-15\,\mathrm{nm}$ in 3D, depending on the specific signal-to-noise in the acquired images. 
The volume methods were applied to a large population of non-genetically perturbed cells and populations of cells with mitotic kinesin-5 molecular motors deleted. 

\section{Overview of feature localization and tracking in biology}
%%%%%%%%%%%%%%%%%%%%%%%%%%%%%%%%%%%%%%%%%%%%%%%%%%%%%%
Feature finding algorithms are designed to automatically localize the intensity centroid of a specific feature in an image. In biological systems, the features of interest are often fluorescently labelled proteins and organelles, with dimensions below the diffraction limit; the intensity distributions to be localized are therefore diffraction limited spots. Feature finding algorithms consist of two main steps to automatically detect and then localize all features of interest in a given image. The first step is an initial detection of local intensity maxima over the entire image, identifying candidate feature locations to pixel resolution. In the second step, the full intensity distribution of the candidate features, which spans many pixels, is used to localize the centroid of each intensity distribution to sub-pixel resolution. 

To extract the real space dynamics of a feature from a time-series of images, tracking algorithms establish the correspondence between features in successive image frames, forming a set of single-particle trajectories. In general, feature correspondence is determined by minimizing the displacement of localized features between frames. 
The ability of tracking algorithms to reconstruct feature trajectories with high fidelity is critically dependent on the ability of the localization algorithm to determine the particle positions accurately.

Among articles that assess the limits of accuracy and precision of particle localization algorithms, 
Cheesum \emph{et al.} \cite{Cheesum_BiophysJ_01} quantitatively compare algorithms for localizing fluorescently labelled objects in two-dimensions, from among  
centre-of-mass algorithms, 
direct Gaussian fitting which exploits the shape of the intensity distribution for a diffraction-limited spot, 
image cross-correlation algorithms, and 
sum-absolute difference methods.
Typically, centre-of-mass and direct Gaussian fitting methods obtain sub-pixel resolution directly. In contrast, both the cross-correlation and sum-absolute difference methods determine the object's position to only pixel resolution. To obtain sub-pixel resolution, the cross-correlation or sum-absolute difference matrices must be interpolated. 
Cheesum \emph{et al.} measured accuracy as the mean difference between the localized position and actual position of a simulated object over a large number of trials and determined, unsurprisingly, the direct Gaussian to be the superior algorithm for localizing point sources. 
%%%%%%%%%%%%%%%%%%%%%%%%%%%%%%%%

Thompson \emph{et al.} examine 
the factors that limit the precision of centroid localization~\cite{Thompson_BiophysJ_02}, for limiting cases where a large number $N$ of signal photons are collected, in which the data is considered photon shot-noise limited; and the limit of small $N$, where the image is considered background noise limited, obtaining the theoretical limiting precision for one spatial dimension, $x$ by considering each of these limits independently and linear interpolation between them~\cite{Thompson_BiophysJ_02}.

In a more general treatment, Ober \emph{et al.} have used a Fisher information matrix to determine the limit of localization precision for single-molecule microscopy \cite{Ober_BiophysJ_04}. 
%The Fisher information matrix plays a central role in the analysis of estimation algorithms. Its inverse is the so-called Cram\'{e}r-Rao lower bound for the variance of any estimation that has a mean equal to the true value; that is, for any \emph{unbiased} estimator. The use of a  Fisher information matrix is a particularly powerful assessment because its prediction depends only on the statistical model of data generation. For a diffraction-limited spot in an image, elements of the information matrix depend only on the PSF and noise models. 
The inverse of the Fisher information matrix is independent of the particular localization procedure. However, importantly, Ober \emph{et al.} assumed the localization procedure would provide an \emph{unbiased} estimate of the object location. This assumption is not justified for all localization routines, as was shown in \cite{Cheesum_BiophysJ_01} and~\cite{Gao_tracking09}. 
The ultimate limit of localization precision, 
as a function of the emission wavelength $\lambda_{em}$, photon emission rate $A$, objective lens numerical aperture $N\!A$, and acquisition time $t$, is $\avg{(\Delta x)^2}=\avg{(\Delta y)^2}=\lambda_{em}/(2\pi \, N\!A \, \sqrt{\gamma \, A \, t})$, 
where $\gamma$ is the optical system efficiency, defined as the fraction of photons leaving the object that reach the detector. This 
is a fundamental limit of localization and does not include the effects of pixelation or noise~\cite{CrockerHoffman-chapter2007}. When these effects are included, the relation is much more complicated, and the limiting precision depends on the particular imaging conditions and detector specifications.

%%%%%%%%%%%%%%%%%%%%%%%%%%%%%%%%%%%%%%%%%%%%%%%%%%

Knowledge of the expected intensity distribution for a diffraction-limited spot is used in feature localization routines to obtain the spatial coordinates of an object with high precision and accuracy. This 
knowledge can also be used to segregate two spots that are closely spaced. Thomann \emph{et al.}  \cite{Thomann_JMicrosp_02} developed a method for automatic detection of diffraction-limited spots separated by less than the Rayleigh limit in three-dimensional image stacks under the key assumption that each spot detected in an image is comprised of a finite number of superimposed PSF's. 
Using simulated data, Thomann \emph{et al.} \cite{Thomann_JMicrosp_02} found that their algorithm can resolve points at sub-Rayleigh separation. The localization accuracy approached the nanometer range for signal-to-noise greater than $12$ dB, and was sub-20 nm for lower signal-to-noise ratios. High accuracy was only maintained for points separated by at least the Rayleigh limit, and depended on the relative brightness of the two points. Spots of equal brightness could be distinguished at a distance of half the Rayleigh limit. For cases where the spots differ in brightness, the resolution limit decreased by up to 50 \%.

Several works address  high fidelity tracking in systems where particle trajectories may overlap~\cite{Serge_NMethods_08, Jaqaman_NMethods_08}, a common problem encountered in using two-dimensional detection to acquire images of three-dimensional objects. Imaging using three dimensional detection of diffraction-limited spots separated by greater than the Rayleigh limit lifts the degeneracy of particle  overlaps and (dis)appearances. 

%%%%%%%%%%%%%%%%%%%%%%%%%%%%%%%%%%%%%%%%%%%%%%%%%%%%%%%%%%%%

 %With current capabilities for carrying out three-dimensional imaging, volumetric measurements that are extrapolations from two-dimensional shape, or approximations to a regular surface of revolution. 
%Robust surface reconstruction methods would also enable locating submicrometer entities at their address in a living cell, an important task in capturing subcellular dynamics. 
Just as the dynamical resolution of point-like objects is limited by the resolution in the feature localization, so also the resolution for volumetric changes of cells or subnuclear regions is limited by the faithfulness of any surface reconstruction being performed.

Many works address the problem of reconstructing the topography of a surface from sample points in three dimensional spaces~\cite{Boissonnat,TurkandLevoy,Bernardini,AmentaBernandKamvysselis,CoconeIntJCompGeomAppl02}. 
Watertight
% (i.e., no holes) 
% returning a well-defined interior and exterior
surface reconstructions, which return a well-defined interior and exterior, are needed for many applications, including computer graphics, computer aided design, medical imaging and solid modeling, but are not guaranteed by many reconstruction methods. 
Basic surface reconstruction techniques can be classified into explicit representations, including   
parametric surfaces and triangulated surfaces, in which all or most of the points are directly interpolated based on structures from computational geometry, such as Delaunay triangulations; 
% connect points into an approximate mesh, and 
and implicit representations, which solve the problem implicitly in 3D space by fitting the point cloud to a basis function, 
%address the surface reconstruction problem 
%by fitting a 3D function to the point samples 
and then extracting the reconstructed surface as an iso-surface of the implicit function.
Methods in the second category differ mainly in the different implicit functions used.
% which solve the problem implicitly in 3D space based on some basis functions. 
Implicit representations are a natural representation for volumetric data: 
%one fits implicit functions on the point cloud, then uses a marching-cube (MC)-like algorithm~\cite{MCproc} to extract the zero-set of the function into a mesh; that is, the iso-surface. 
%, defining a sort of potential field in the space surrounding the target surface, allowing extraction of a iso-surface. 
%Solid objects can have the surface located by isosurface methods, which are notoriously sensitive to the threshold chosen. 
%An implicit method of some type plus marching cubes is currently the most popular algorithm for isosurface extraction from medical scans (CT, MRI) or other volumetric data~\cite{MCproc}. 
the marching-cube (MC)-like algorithm~\cite{MCproc}  is currently the most popular algorithm for isosurface extraction from medical scans (CT, MRI) or other volumetric data~\cite{MCproc}. MC has been applied to subcellular~\cite{GuilakRatcliffe1995,Guilak1995} and cellular~\cite{Zonal-chondocyte-3Dvolume,compressibililty-chondrocyte-2011} volumetric data using commercially-available software applied to three-dimensional image data obtained by confocal microscopy. 
% In general isosurface methods are notoriously sensitive to the threshold chosen. 
MC produces the desired watertight, closed surface and is easy to parallelize, but requires uniform (over-)sampling, produces degenerate triangles (requiring post-processing re-meshing), and does not preserve features. While the output mesh that MC generates is adequate for visualization purposes, it is far from being suitable for surface or volume quantification in situations of local undersampling due to high curvature. 

In practice, the data may sample only part of a surface densely, and may suffer from inadequate sampling in other, local  regions, either due to non-perfect homogeneous distribution of a fluorescent reporter, or to local high curvature of those regions relative to the imaging (pixel) sampling rate.  
For example, in budding yeast, due to the small cell size and the even smaller incipient daughter cell (bud) size, there are local regions in which the curvature may be high relative to the sampling, even when sampled at the highest microscope magnification. 
Existing surface reconstruction algorithms face difficulty  if the data contains local undersampling, and will yield 
holes or artifacts under these conditions. We envision many scenarios where local undersampling is present.  

The tight Cocone method generates a surface representation given a set of sample points, 
% handling undersampling, specifically, 
suppressing undesirable triangles and repairing and filling the holes in the surface~\cite{acmprocsolidmodeling}.
% , producing water-tight surfaces from an approximate mesh first generated from Delauney ``tetrahedralization''. 
% It provides a solution that works in practical conditions of undersampling and on complex surfaces. 
It provides a water-tight reconstruction even when the sampling bounds required by other algorithms are not met. 
For surfaces that are complex, i.e. that are not single compartments, the tight Cocone algorithm 
% will only reconstruct surfaces that separate the interior volume from a point at infinity, 
will reconstruct the full surface. 
% We add an additional step of decomposing the space. 
By decomposing the space and re-running the algorithm, we are able to build more complete volumetric representation of the  dividing yeast cells, producing watertight reconstructions of each compartment.
%
%output of a fit to two ellipsoids, itself run on the initial complex surface. Employing the algorithm in this way allows us to build more complete volumetric representation of our dividing cells than a simple space carving approach, producing watertight reconstructions of each compartment.
% , effectively decomposing the complex surface into single convex hulls. 

In this  work, feature finding and tracking in living cells was developed together with surface reconstruction and volume determination. 
%to obtain coordinates relative to important topographic boundaries. 
The algorithms are semi-automated to maximize the number of cells that can be analyzed with a minimal amount of subjective manual input. 
These new methods were applied to investigate budding yeast mitotic spindle dynamics at a high spatial resolution over the entire cell cycle, and the compartment volumes of dividing cells. 

\section{Automated localization of mitotic spindle poles, and robust surface landmarks and volumes of challengingly small cells}
To analyze the spatial location of mitotic spindle poles in a reference frame intrinsic to the dividing {\em S. cerevisiae} cell, we determined the 3D position of the poles 
%relative to 
and, in parallel, the cell bounding surface, the plane of cell division (neck plane), and  neck plane center, as landmarks.  
To fluorescently tag spindle poles, we used cells in which the chromosomal copy of Spc42 was fused to the red fluorescent tandem-dimer Tomato. We additionally used cells in which the chromosomal copy of the G-protein Gpa1 was fused to GFP, to visualize the cell surface in the same cells. We imaged asynchronous live-cell populations in three dimensions, yielding images containing up to $\sim20$ cells each.  

\begin{figure}[ht]
  \begin{center}
 \includegraphics[width=0.5\textwidth]{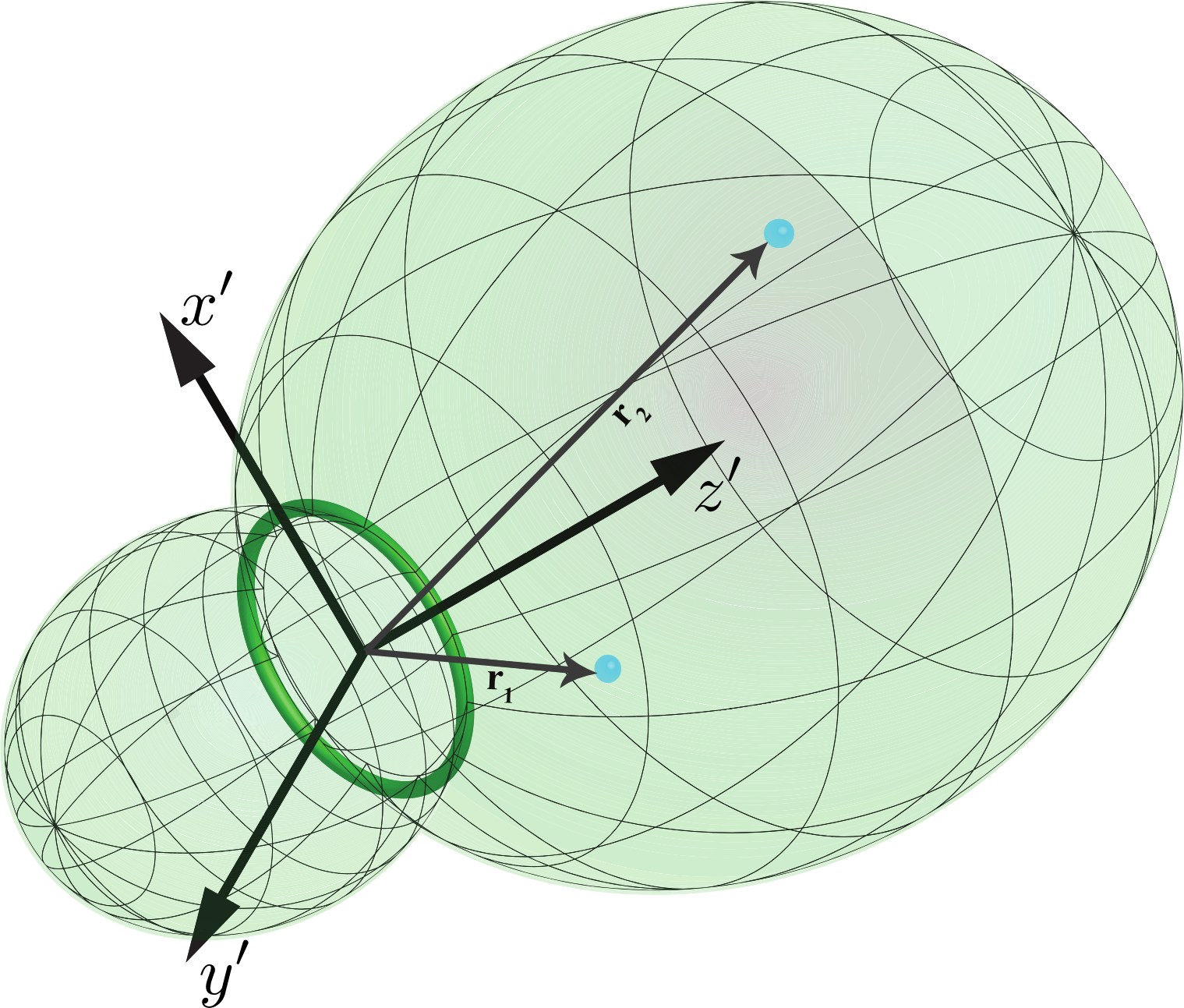}
    \caption{Illustration of the coordinate system for spindle poles as defined by the cell periphery for budded cells. $x^{\prime}$ and $y^{\prime}$ lie in the plane of separating the mother and bud cavity during cell division, called the {\em bud neck} in budding yeast. The $z^{\prime}$ coordinate in this definition is positive in the mother cavity and negative in the bud.}
  \label{fig:img_sim:cell_coord}
\end{center} 
\end{figure}

Experiments were performed at a temperature of 25 \degree C. At this temperature the budding yeast cell cycle is $\sim\!2$ hours. 
Photobleaching restricts the window of time that a single cell can be observed at several nanometer resolution. Thus, to capture spindle dynamics over the entire cell division, populations of unsynchronized cells were observed, with each cell in a field-of-view acting as a sample of a specific temporal window of the cell cycle. 
The spindle poles were  labelled with one fluorescent protein and cell periphery was labelled with a different fluorescent protein. 
The spindle pole trajectories and mother and bud cavity surfaces were reconstructed from fluorescence data as described below.

The final result of pole and surface finding is an ensemble of spindle pole trajectories, with the capability for each trajectory to be described in a coordinate system relevant to  cell division for that cell. The geometry resulting from such an analysis for a single cell is plotted in Fig.~\ref{fig:img_sim:cell_coord}. In the next section, the algorithms used for extracting spindle pole dynamics and cell surface topography are described. 
% This specific implementation was originally developed in the Kilfoil lab by Dr. Vincent Pelletier and further extended in this  work. 
All implementations were written in MATLAB (Mathworks, Natick, MA).

\subsection{Three-dimensional tracking of point-like features}\label{sec:FF:3dpoint}
\begin{figure}[ht]
  \begin{center}
\subfigure{
  \includegraphics[width=0.32\textwidth]{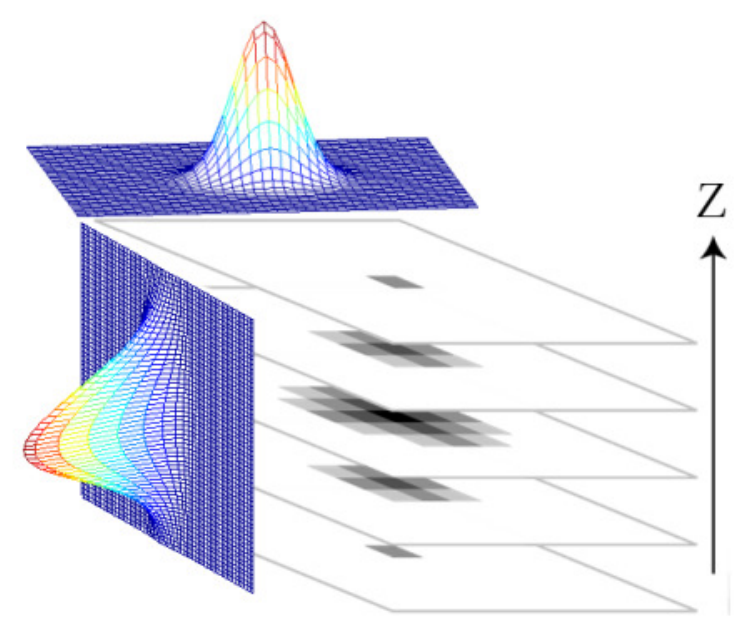}
}
\subfigure{
   \includegraphics[width=0.12\textwidth]{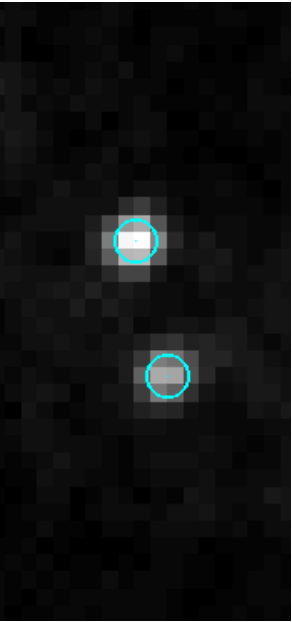}
}
    \caption{The general principle of point localization. The intensity distribution of a diffraction-limited spot is fit to a three-dimensional Gaussian function. Localized spindle poles, superimposed on the signal projected  into the plane of imaging, are shown on the right.}
  \label{fig:FF:3D_ff_principle}
\end{center} 
\end{figure}

As described above, knowledge of the expected intensity distribution for a diffraction-limited spot can be used to automatically localize fluorescently-labelled objects that are smaller than the diffraction-limit. The general principle of the localization routine is depicted in Fig.~\ref{fig:FF:3D_ff_principle}. By fitting the full intensity distribution of the object to a three-dimensional Gaussian function, the centroid of the spot can be determined to sub-pixel resolution in all three dimensions.
To construct spindle pole trajectories from a time series of confocal fluorescence image stacks, an automated three-dimensional feature-finding and tracking algorithm was used. The algorithm locates the centroids of the spindle pole intensity distributions, termed features, to sub-pixel resolution in 3D in a series of successive confocal image stacks acquired at controlled time intervals. It was originally developed as center-of-mass fitting for applications in colloidal systems at 3D sub pixel resolution~\cite{Gao_tracking09}, and adapted for application in living cells using Gaussian fitting by two of the authors. 
The tracking algorithm links localized features in sequential image stacks and outputs a time series of coordinates for each feature. The resolution of the three-dimensional localization depends on the signal-to-noise ratio of the images, 
as will be discussed in detail below. For this study, the localization accuracy is between  $9$ and $12$ nm in 3D.

The automated feature finding consists of three main steps: filtering, initial estimation of feature positions, and refinement of position estimates. To suppress image artifacts and noise, which exhibit high spatial frequencies (small length scale), the images are initially filtered using a three-dimensional band-pass Gaussian kernel. By choosing a spatial cut-off for the filter kernel close to the characteristic size of a diffraction-limited spot, high frequency noise is attenuated to a level far below that of true features. Once the image has been filtered, pixels that are local intensity maxima can be used as initial estimates for feature positions. To further distinguish real features from noise, the integrated intensity is calculated for a sub-volume surrounding each of the local maxima. Features for which the integrated intensity is below a cut-off value are excluded from further processing. At this point, the feature position estimates have a spatial resolution equal to the pixel size, typically 180 nm in $\hat{x}$ and $\hat{y}$ and 300 nm in $\hat{z}$.  To refine the feature position, a three-dimensional Gaussian function is fit to the intensity distribution within the sub-volume surrounding the initial estimate, using non-linear least squares. The final sub-pixel estimation of the feature location is obtained by a second iteration of Gaussian fitting. The second iteration involves first shifting the pixel-spaced sub-volume so that it is distributed around the centroid of the initial Gaussian fit~\cite{crockergrier96,Gao_tracking09}, followed by a second least squares Gaussian function fit, using the outputs of the initial fit as parameter estimates for the second fit. 

The band-pass length scale, the dimensions of the sub-volume used for fitting, the integrated intensity cut-off value, and a characteristic size of the object are used to define a region around the candidate feature in which a second feature cannot be located, and are defined by the user. The last parameter is adjusted to allow or reject any overlap between two or more features of interest. These four parameters are set to optimize the fidelity of feature localization across a large number of cells contained in a single field-of-view. Since photobleaching degrades the signal of a given feature as it is observed over time, the integrated intensity cut-off can be automatically reduced by a user-defined constant factor for each timepoint in the series of confocal stacks.

The tracking algorithm links localized features in sequential image stacks and outputs a time series of coordinates for each feature. 
The algorithm used here is based on Crocker \emph{et al.}~\cite{crockergrier96} and implemented in Matlab~\cite{Pelletier}. 
Feature coordinates are linked in time by minimizing the total displacement of each of the identified features between successive frames. The search for corresponding features in consecutive frames is constrained to a maximum likely displacement over the timescale between frames. 
Accurate and robust tracking is obtained by tailoring the timescale of image acquisition and the cut-off distance to the particular dynamics of the features being tracked. 
High fidelity tracking is obtained by applying successive iterations of the tracking algorithm to each data set~\cite{Gao_tracking09}. This method 
ensures that the majority of trajectories over a cell population are captured, and results in data sets biased for trajectories that persist to long times.

\subsection{Automated watertight surface reconstruction robust to local undersampling}\label{sec:ff:3dsurf} 

The cell periphery is visualized using fluorescent reporters that localize to the cell cortex. Using the procedure we developed, if the reporter is distributed uniformly over the entire surface, the fluorescence intensity distribution in a confocal image stack can be used to reconstruct the cell surface. Automated surface reconstruction is carried out on stacks of images obtained using reporters that label the periphery of living cells, provided the resolution in $z$ is sufficient and the label (dye or fluorescent reporter protein) distributes uniformly around the cell periphery. Key features of the surface shape that are important to cell division, such as the bud neck in budding yeast, may then be extracted. 

The plane of cell division in budding yeast is called the bud neck: in any dividing cell it is characterized by local deformation of the bounding surface to create two cavities out of one. In addition to the final topology change, local geometry is also modified smoothly during the process of cell division.  Using the mother and bud cavity fits for each cell, the neck plane position and orientation can be calculated using the methods described here.

The automated surface reconstruction procedure we developed consists of five main steps: filtering and thinning, initial estimation of closed bounding surface, refinement of surface estimates by decomposition of the space, calculation of volume of each convex cavity (here, mother and bud cavity of mitotic cells), and, in mitotic cells, determination of position, orientation, and size of  the neck separating the mother and bud cavities, which provides a basis coordinate system for cell division unique to each cell. 

\subsubsection{Filtering and thinning: }
The first step of surface feature finding begins with spatial filtering of the image stack, with the characteristic size parameter of the 3D band pass algorithm corresponding to the surface thickness. Using a first estimate of the thickness, one can iterate through different values until the surface and a reasonable cavity size are reproduced with optimum fidelity. 
The thresholding step must use a threshold value that maximizes the amount of surface included, while minimizing artefacts such as protrusions, small clusters away from the surface, and non-faithful connection between opposite sides of a cavity. % Small clusters not connected to the surface can be removed by setting a minimum size of clusters of connected voxels: 
Typically, the true surface data is comprised of thousands of connected voxels, whereas small noise clusters may span only a few hundred voxels or less, providing an effective means for the identification and removal of small noise clusters.

%%%%%% THUS CAN MINIMIZE RELIANCE ON THRESHOLDING
%
%While the cell surface may be inherently variable, the surface data of interest will typically comprise thousands of voxels, whereas small noise clusters may span only a few hundred voxels or less, and can thus be distinguished and removed readily. 

The thresholded image stack is then sliced in $(x,y)$-plane images, each of which is ``thinned'' to a minimal collection of points that preserves holes and general shape, to allow for a full surface reconstruction. Thinning was used instead of skeletonization which is less suited for minimizing the surface data as it  sporadically produces branches. 
% since the algorithm implemented in the image processing toolbox in Matlab requires 2D images. This 2D 
The thinning algorithm  transforms an entity within the image into a single pixel thickness entity~\cite{thinning}.
The algorithm yields good results on slices near the equator of the surface to be analyzed, where it reduces a thick shell to a 1 pixel-wide shell. An example on one focal plane of data acquired from the surface reporter imaging in budding yeast, before and after the thinning step, are shown in Fig.~\ref{fig:meth_surf3} (a-b). On nearly-flat curved surfaces, which are manifest in slices through the top or bottom of a cell, 
the thinning algorithm has the undesired effect of reducing a uniform disk to a single or branching lines. 
By additionally slicing the image stack into sets of $(x,z)$ and $(y,z)$ planes, and processing these sets of planes with the thinning algorithm, we eliminate these artifacts specific to the slicing direction. In these new slices, ``top'' and ``bottom'' caps are instead part of equatorial profiles. 
The result produces good performance, defining the top and bottom of the cell as a collection of points in a contour. 
The estimated minimal set of points describing the surface is then constructed by merging the results from the $(x,y)$, 
$(x,z)$ and $(y,z)$ slice directions. We assemble the final cloud of points by  requiring that for a point in space to be retained, it must appear in at least two of the slicing directions. 

The resulting 3D thinned version of the thresholded image stack is composed of a cloud of points centered approximately in the middle of the original thick surface of the image stack. 
Again small disconnected clusters of points can be removed from the thinned surface by requiring a certain minimum size of connected points. 
The output of this discrete cloud of points estimation step is used as input to the Voronoi-based algorithm to obtain useful surface representation, which can then be used to quantify volume. 

%obtain closed surface
\subsubsection{Initial estimation of closed bounding surface: }
\begin{figure}
   \begin{center}
     \includegraphics[width=0.45\textwidth]{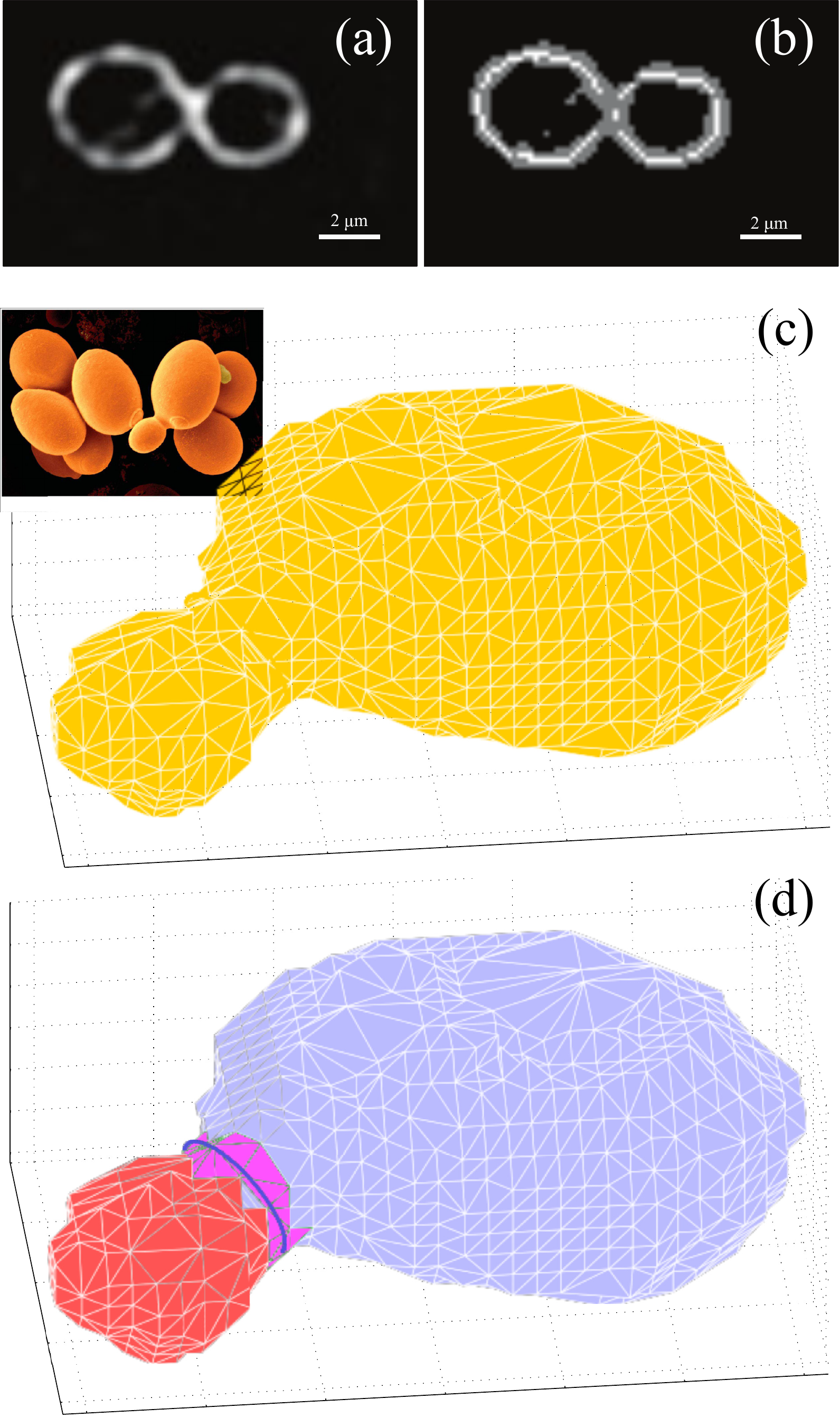}
      \caption{Demonstration of the algorithm on practical data. (a-b) Filtered and thinned surface profile. (a), a central confocal plane of the processed surface data after filtering. (b), light grey pixels are the data after applying the intensity threshold. White pixels are the result of thinning.    (c) First stage of the algorithm which generates a full water-tight surface with sufficient homeomorophic representation 
  of the intended surface. The output has the exact topology and approximate geometry of the sampled surface. Inset shows SEM image of {\em S. cerevisiae} for comparison. (d) Result of the second stage of the algorithm which generates automated segregation of the reconstructed surface into mother cavity (violet), bud cavity (red), and neck plane (magenta). The neck surface is fitted to a tilted ellipse (dark blue) to determine neck plane location and orientation.}
      \label{fig:meth_surf3}
      \hspace{0.06in}
   \end{center}
\end{figure}

To define the volume, the scattered point samples of the surface need to be turned into a proper watertight surface mesh. 
In cases where the signal is strong and uniform, an accurate, arbitrarily-shaped tessellated surface can be obtained via a full surface reconstruction. We achieve this using the Tight Cocone software~\cite{CoconeIntJCompGeomAppl02,acmprocsolidmodeling} applied to the minimized discrete surface points. The Tight Cocone algorithm takes a point distribution in 3D and attempts to reconstruct a water-tight surface described by those points. 
The algorithm is robust to local undersampling as intended, but as it is sensitive to noise,  we supply the filtered data. 
The output is a list of triangles that comprise the reconstructed surface. As a proof-of-concept example, the result on one budding yeast cell for which the surface is thus tessellated is shown in Fig.~\ref{fig:meth_surf3}. The determined surface is remarkably faithful to the actual cell's topography, reproducing bumps, curves, and the typical smooth curvature transitions %produced by septin ring complexes 
at the bud neck. 

%This method relies of the locality of undersampling.  
%Also, the method presented in the paper does not reconstruct internal voids. 

\subsubsection{Decomposition of complex surface:}
% calc volume
From a reconstructed surface, quantities such as volume may then be computed. If the shape is a single convex hull, the volume can be computed simply, by using a well-known algorithm. 
If the complete surface is not a single convex hull, this approach will not robustly provide the volume.
For example, in budding yeast, the bud cavity can protrude significantly out of the mother, with a concave neck region linking the two cavities. However, individually, the mother and  bud cavities are convex. 
% and we adapted this approach for this application. 
We effectively decompose the complex multiple-hull surface into individual convex hulls, for each of which  compartment the volume can be determined.

In principle,  the negative curvature of the neck region could be used to obtain the orientation and location of the plane separating the mother and bud cavities.  In practice, even though the periphery is well-labeled so that the whole surface can be reconstructed, undersampling prevents determination of the local curvature everywhere. 
Moreover, in small budded cells the neck curvature may only just change sign at the transition region from mother to bud.  
Instead, we exploit other information to uniquely identify the tessellated surface data in this region between the two cavities. 
In our dividing cells, we subdivide the surface into sections by performing a fit of two ellipsoids to the minimized surface points.  
% The result of the initial fit is two ellipsoids that are reasonable estimates of the mother and bud cavities. 
For each vertex of the closed bounding surface, we then compute the shortest distance to the fitted ellipsoids. The computed 
proximity is used to segregate the vertices into two subsets: those assigned to the mother cavity, and those assigned to the bud cavity. % In a second iteration, an ellipsoid is fit to each of these subsets independently, using the output of the first iteration for initial parameter estimates. 
%This fitting can alternatively be performed on data that has not been thinned, with each surface point weighted in the fitting algorithm by its intensity. 
% Each point is identified as belonging to either the mother cavity or the bud cavity 

%, segregate all of the points into mother and bud sets. 
After this assignment step, we perform the water-tight surface reconstruction 
on each segregated point set. This provides individual convex surfaces which are used for  
robust calculation of volumes for each of the two cavities.
%Remaining volume unaccounted for is bounded by the two closed convex cavity surfaces, and by 
Following this step, those triangles in the original full surface reconstruction having at least one vertex belonging to each of the mother and bud sets are assigned to the neck region, and are used to reconstruct the surface of the neck region. 
All of these steps are carried out for each dividing cell.  
%This can allow for calculation of the  neck volume, if desired.
% Among the original surface polyhedra, 
%Due to the small number of points thus defined, and the convex nature of the mother and bud cavities, this region's surface will be reconstructed less reliably. While the calculated neck volume will include a portion of the mother or bud volumes, this approach will still give a robust result and provides a good approximation of the size and volume of the neck region. 
%Clearly the most accurate fit will be found when a uniformly distributed set of points populate the neck surface. 

\subsubsection{Calculation of volume of each convex cavity: }
The volume of a convex hull can be computed simply, as follows.
Considering any point on the surface of or inside the volume, the volume is calculated as the sum of all pyramids whose base is one of the triangles forming the surface and whose summit is that point. These pyramids constitute a complete set,  containing the volume enclosed by the surface, and the sum of their volumes is the total volume of the convex hull. 

\subsubsection{Determination of position, orientation, and size of bud neck: }
From the minimized points forming the neck surface, we extract geometric characteristics of the neck by fitting a circle (or ellipse) to these points.  This yields the neck plane location (including the center); circumference; and, via its normal, orientation with respect to the imaging frame of reference. Knowledge of this plane provides a reference frame for cell division unique to each cell. As can be seen in Fig.~\ref{fig:meth_surf3} (d), this automated procedure results in 
faithful description of the orientation and position of the neck plane between the mother and bud cavities. The results obtained show the suitability of the method for a correct representation of the target object.
The trajectories of other point-like features obtained simultaneously with the surface in each cell can now be transformed into their cell-unique coordinate system, and therefore examined relative to the neck origin.  

Together, the spindle position and surface mapping could additionally be used in tandem to define a wealth of physical observables. Measuring the dynamics of the mitotic spindle in a coordinate space defined by the plane of cell division, over a population of cells, would reveal dynamical and shape cues related to stages of the cell cycle not quantifiable precisely by other methods. 
Specifically, spindle orientation, and translocation with respect to the neck plane, could be obtained automatically.
The orientation of the normal vector to the neck plane, $\vec{z^{\prime}}$ is determined from the fitting as described above,  and its sign can be assigned as desired: we here define $\hat{z^{\prime}}$ as extending into the mother cavity. 
A spindle vector $\vec{s}$ may be defined as the vector spanning from the pole proximal to the bud to the pole distal to the bud. Spindle orientation is then parametrized by the angle $\psi$ between $\vec{s}$ and $\vec{z^{\prime}}$, given by $\vec{s}\cdot \vec{z^{\prime}}=|\vec{s}||\vec{z^{\prime}}|\cos \psi$, where $0 < \psi < 90^{\circ}$. 
For each  cell,
the spindle orientation and its angular excursions may be calculated over the observation window. 

The distance of the spindle poles from the neck plane may be computed by constructing the vectors $\vec{r}_{xn}$, representing the displacement of each SPB located at point $x_n$ from the neck centre at point $x_{0}$. The distance, given by $d=\vec{z^{\prime}}\cdot\vec{r}_{xn}$, is positive for all $x_n$ in the mother cavity, and negative for all $x_n$ in the bud cavity, using our definition of the $\vec{z^{\prime}}$ direction.

Accurate segregation of aberrant spindle positioning behavior from that of normal cells 
%during spindle breakdown 
would require knowledge of the spindle size and position relative to the mother and bud cavities.
%, achievable in cells  in SC medium with both poles and surfaces labelled. 
Such studies, achievable at high resolution in cells in SC medium 
with both poles and surfaces labelled, will be enabled by these tools. 

The ability to calculate compartment volume might have other uses. 
% Dealing with high curvature or irregularly shaped compartments.  
Not all volumes one wants to quantify are conveniently regular. 
Chromosome territories, nucleoli, and Golgi bodies are all examples of compartments with complex distributions or irregular outlines, occupying in higher eukaryotes approximately the same volume as the yeast cells under study here, a few tens of femtoliters~\cite{Voronoi-chromsm-volumes-1995}.
Methods to study such compartments in a statistically meaningful manner have been limited and are generally complicated by the fact that estimating the volume of such exclusion zones requires calculating the volume of irregular domains.

%should you describe the microscope here?? 
%probably good to have schematic of object image and pixel spaces here. 
\section{Results and discussion}\label{sec:res}
\subsection{Characterization of the imaging system}
The accuracy of feature finding and tracking in cells is influenced by many factors. Signal-to-noise must be maximized with well defined intensity distributions for each feature, spanning multiple voxels so that the Nyquist criterion is satisfied. The hardware, signal-to-noise, and sampling requirements limit the achievable temporal resolution. Investigation of fast dynamics in the cell requires some trade-off between temporal and spatial resolution. A detailed assessment of the factors influencing the performance of localization algorithms to determine the limits of the spatial resolution greatly aids in the design and interpretation of feature finding experiments.

A simulation of diffraction-limited spots that is relevant to the experimental conditions requires knowledge of the true PSF of the optical system used for the experiment. The theoretical description of a confocal PSF is only relevant in the ideal aberration-free scenario, and it is necessary to measure the real PSF of the microscope. Measurement of the PSF will reveal any aberrations that may be present in the optical system.

The image of an object that is smaller than the diffraction-limit closely approximates the true PSF of the microscope. In an optical system, a diffraction-limited object acts as a low-pass spatial filter with a high cut-off frequency relative to that of the OTF. The cut-off frequency increases as the inverse of the object size. The true PSF is the image of an infinitesimal object, i.e.\ one with an infinite bandwidth. However, in practical imaging conditions, background and noise signals often mask any contribution of high frequency components to the PSF, so that the effect of a small but non-zero object size on the PSF is negligible. Therefore subresolution fluorescent beads may be used to measure the PSF. Measurement of the PSF for our optical system is described in the supplementary materials~\cite{suppmat-exptdetails}.

To simulate all aspects of image acquisition realistically, the background or bias signal and the readout noise from the CCD camera were also measured, as also described in~\cite{suppmat-exptdetails}.

\subsection{Performance test on synthetic data}\label{sec:img_sim:poles} 
% Simulation of spindle poles
In order to determine the accuracy and precision associated with the feature localization algorithms used in this hybrid method, the localization algorithms were applied to computer generated confocal images of spindle pole bodies. Our procedure for generating these simulated images, using parameters drawn directly from our hardware, signal-to-noise, and 3D sampling characteristics, is described in~\cite{suppmat-exptdetails}.

For each simulation run, an image was generated at a given signal level and with a given displacement between the two diffraction-limited spots in a prescribed direction in three-dimensional space. Noise characteristic of the yeast medium and camera (as measured;~\cite{suppmat-exptdetails}) was added to the image, and the feature localization algorithm was applied. At each signal level and spot separation tested, 500 iterations of feature finding were performed with a new noise profile generated for each iteration. All simulations were performed in MATLAB. Feature localization parameters were set to the values used for live cell trials. The feature coordinates, as determined by the localization algorithm, were then compared to the real coordinates to obtain an estimate of the localization error over a range of signal-to-noise values and spot separations. Knowledge of the true spot locations allows for an estimate of both precision and accuracy of a given measurement. 
\begin{figure}[h!]
\centering
  \includegraphics[width=0.48\textwidth]{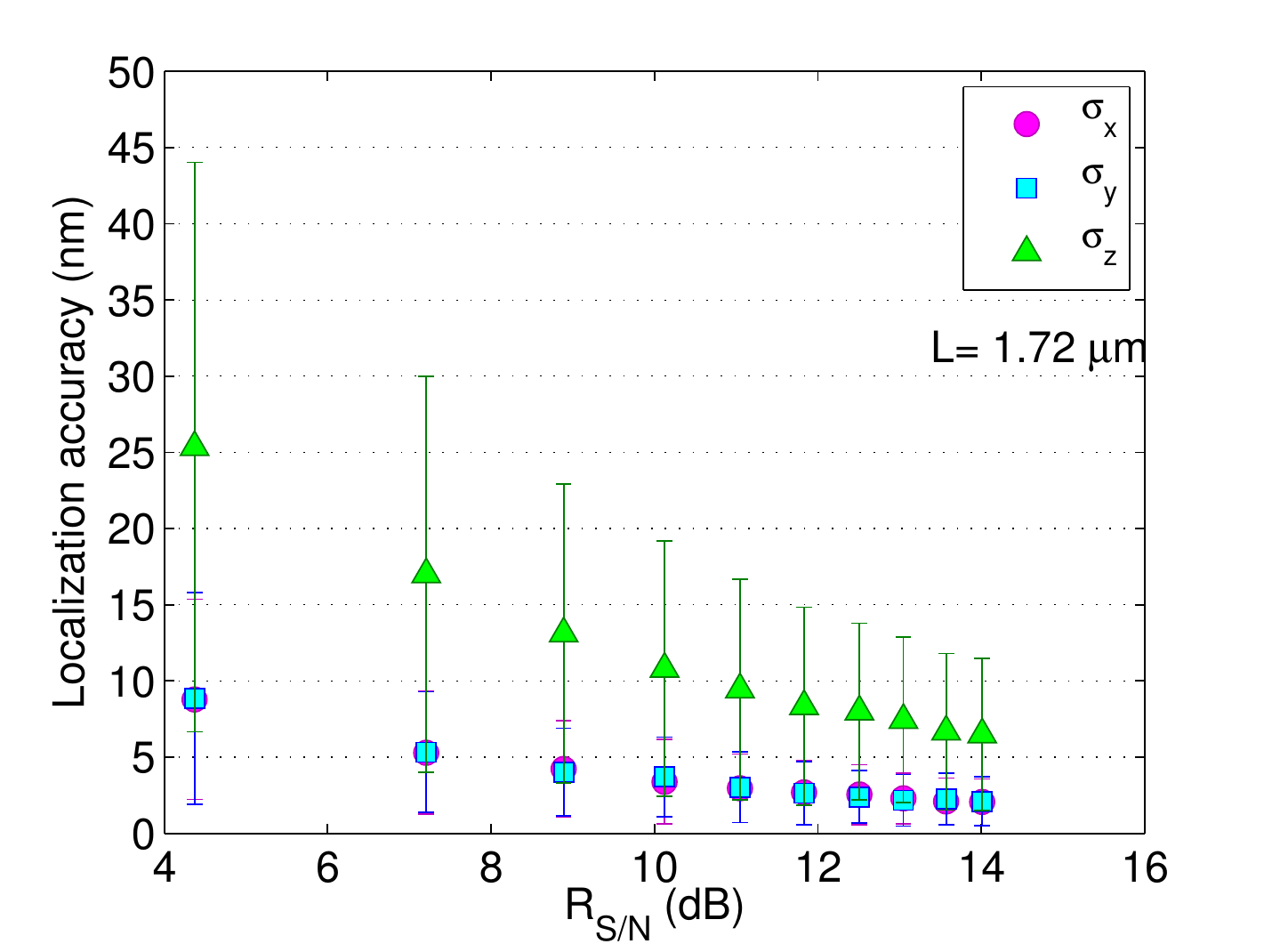}
   \caption{Point localization error versus signal-to-noise for a separation typical of the metaphase spindle sampled over a range of spindle orientations. 
   Each data point represents the mean localization error over 500 iterations. The error bars are the variance, which represents the precision of the measurement.}
  \label{fig:img_sim:Error_vs_RSN} 
\end{figure}
The deviation between the actual position and the localized position was derived in this way. The result for a point-to-point displacement typical of a metaphase spindle, over a range of signal-to-noise levels,  is shown in Fig.~ \ref{fig:img_sim:Error_vs_RSN}. Anaphase spindles have greater separation and even less vector component along  the $\hat{z}$ direction, so  suffer less from localization error than than do  metaphase spindles. 

% \subsubsection{Influence of overlapping of features}
Budding yeast cells are small, rendering subcellular quantitation challenging.  Nevertheless, the spindle poles in these cells are separated by $1.5-2\,\mu\mathrm{m}$ in the early phase of cell division (metaphase), and by larger distance later in the cell cycle. Thus the metaphase spindle is the more stringent test of true subpixel and aberration-free resolving of the spindle pole features from one another.  In our applications studied here, metaphase and anaphase, the spindle pole PSFs in our optical system do not overlap. Using the simulations we had developed, we tested the effect of overlap at distances of the same order as, and smaller than, our metaphase spindle length, to validate that effects of overlapping of features need not be taken into account at these distances, and determine at what separations such effects play a role.  

Interestingly, we found that when filtering is used, the filtering itself is a source of small overestimation of the separations at distances $<\sim 2 \sigma_{PSF}$. 
For such separations, spatial band-pass filtering attenuates the summed intensity distribution in the region between the two points, 
and forces the least-squares fitting routine to localize the centroid of each intensity distribution away from the overlap region, resulting in overestimation of the point-to-point separation. 
As overlap increases, the centroids of each feature are forced back towards the overlap region, initially decreasing the estimated separation, until significant overlap has occurred at $\sim\sigma_{PSF}$, and the two spots are difficult to distinguish (see Figs.~\ref{fig:img_sim:2poinIntDist},~\ref{fig:img_sim:Msep_vs_Asep}).

For automated feature detection with live cell image data, spatial band-pass filtering is often necessary. Real data is subject to many sources of noise, and accurate automated segregation of the real features of interest from noise is often not possible without spatial filtering. Therefore, although our feature separations here well exceed this limit, the systematic bias demonstrated here will always contribute to the localization error for closely-spaced features. Although small, this effect must be considered in the interpretation of the results of a feature finding experiment under such conditions.  

Propagation of localization error in the separation of two points was studied in order to determine its role, if any, in the observable $L(t)$, the spindle pole-pole separation~\cite{suppmat-exptdetails}. Comparison of measured error with propagated error for point-to-point separation showed that for large separations, 
%one our case for $L\ge1.4\,\mu\mathrm{m}$, 
the covariant terms are negligible (see Fig.~\ref{fig:img_sim:ErrorPropCompare}). As the two features approach one another to within less than $1.3\,\mu\mathrm{m}$, the covariant terms become significant and they must be included for an accurate estimate of the error. 

We determined that these two effects must be taken into account when the distance between two spots is such that their point spread functions overlap~\cite{Thomann_JMicrosp_02}. %(REFS Danseur, others).
In our observable, budding yeast spindle length in metaphase and anaphase, the spindle pole-pole distances we are measuring always exceed these limits. 
%, but these effects would need to be taken into account were these tools to be applied to real experimental data scenarios where the spots overlap.  

% MAY USE BUT HAVE ALREADY SAID
%These simulations of diffraction-limited spots and cell periphery presented in this section are of value beyond a means to obtain an error estimate. These methods can be extended to any object geometry and used to investigate the effects of the majority of experimental parameters associated with image acquisition, an invaluable aid for both the design of new experiments and the interpretation of results.

% spindle and stages of cell

\subsection{Performance on biological data}
We used budding yeast as a model system in which to study mitotic spindle dynamics. 

\subsubsection{Microscopy and sample preparation}\label{sec:results:microscope}
Live unsynchronized cell populations expressing tdTomato-tagged spc42, a protein in the spindle pole body (SPB) protein complex, and the surface reporter gpa1-EGFP, were imaged by confocal fluorescence microscopy. SPB's labelled with a red fluorescent protein variant (tdTomato29~\cite{Shaner-tdTomato}) at the protein spc42 are point-like features which were localized and tracked using the methodology described above. Surface data was analyzed using the surface fitting described above.

\begin{figure}[ht]
  \begin{center}
 \includegraphics[width=0.48\textwidth]{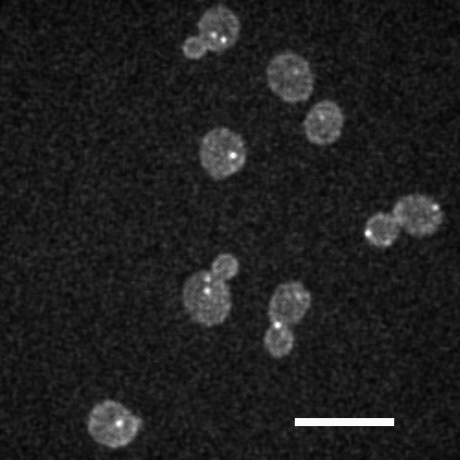}
    \caption{Superimposed confocal planes of budding yeast cells expressing surface (GFP-Gpa1) and spindle pole body (tdTomato-spc42) reporters. Scale bar is $10\,\mu\mathrm{m}$.}
  \label{fig:results:SC_WTtraj}
  \vspace{-0.8cm}
\end{center} 
\end{figure}

All yeast strains used were derived from strain BY4741~\cite{Boekeyeaststrain}. 
Media used for yeast culture are described in~\cite{CSHL_manual}. 
Yeast genetic manipulations were carried out as described in \cite{guthrie02}. 
For microscopy, yeast strains were incubated in SC at 25\degree C until early logarithmic phase ($0.3 \pm 0.05$ normalized optical density at $600$ nm). To obtain the desired cell number density for imaging, 1 ml of suspended cells were pelleted for 15 s at $500\times g$, washed twice with $1\,\rm{ml}$ of lactate medium, then re-suspended by gentle mixing 100 times with 
a pipet, and $8\,\mu\mathrm{L}$ of culture was pipetted onto a SC-infused agar pad on a slide and covered with a glass coverslip.

Cells were imaged at 30\,$^{\circ}$C in stacks of 21 focal planes spaced $0.3\,\mu\rm{m}$ apart, spanning the entirety of the cells in the axial direction, at 10 s per stack, using a custom multi-beam scanner ($512\times512$ pinholes) confocal microscope system (VisiTech) mounted on a Leica DMIRB equipped with a Nanodrive piezo stage (ASI) and $491\,\mathrm{nm}$ and $532\,\mathrm{nm}$ solid state lasers controlled by an AOTF. Images were acquired with a EM-CCD camera (Hamamatsu) using a 63X 1.4 NA objective.  
From the images, we computed the positions of the spindle pole bodies at three-dimensional sub-pixel resolution~\cite{Gao_tracking09} by fitting the intensity distribution to a three-dimensional Gaussian function (Fig.~\ref{fig:FF:3D_ff_principle} (b)), obtaining a spatial resolution of $<10\,\mathrm{nm}$ in 3D for each pole. All analysis was carried out using custom scripts in MATLAB (MathWorks). 

For imaging surfaces, yeast strain of exactly the same genotype with a G-protein labeled as a GFP clone was obtained (Life Technologies, \cite{WeissmanOShea-GFPlibrary}).  The  G-protein, Gpa1-EGFP, localizes to the cell cortex and labels the whole cell periphery~\cite{Bardwell-Gproteins,Dohlman-Gproteins}.  
For all analysis, data was collected from cell populations representing four independently-derived yeast strains. 
For combining of position data from the two channels, chromatic shift between the $\lambda_{ex}=491\,\mathrm{nm}$ and $\lambda_{ex}=532\,\mathrm{nm}$  excitation channels was measured and corrected for prior to combining results (see~\cite{suppmat-exptdetails}). 

\subsubsection{Effect of imaging media on spindle dynamics} \label{ssec:media}
We discovered that for yeast cells imaged in lactate medium, exposure to laser irradiation for image stack collection at 5 second intervals results in a phototoxic effect that is not alleviated by increasing the quantum efficiency of the fluorescent reporters. The effect causes cells to arrest at the transition from metaphase to anaphase.
This effect was confirmed by observing cells in different media types, and varying exposure times for prolonged periods under sealed coverslips. 
 Literature-reported values for the rapid phase of anaphase B in budding yeast demonstrate that this rapid phase is a slowly-varying process compared to the image stack acquisition rate we were using. Therefore, our acquisition scheme should capture the transition from pre-anaphase to anaphase.  % much slower than the sampling rate, and 
Specifically, in observation of an asynchronous budding yeast cell population over a finite 20-minute observation window, representing $\sim$1/5th of the cell cycle, the transition should be observed for $\sim$30 cells in an analyzed population of 160 total cells. Initially imaging in lactate medium, we observed a major discrepancy from the expected biological phenotype: not a single cell traversed to anaphase while under observation. 

To investigate this and determine carefully in what regime the cells can be observed without any photon-absorption-induced perturbation, wild-type yeast strains expressing the spindle pole reporter spc42-tdTomato were prepared for microscopy as discussed above using lactate medium. A small population of asynchronous cells were imaged under a sealed coverslip for 30 minutes at 20 second exposure intervals. Six different fields of view were examined with approximately ten cells per field of view. No qualitative evidence of anaphase spindle elongation was observable in any of the cells. Another population of cells was cultured, washed and imaged in a minimal medium, synthetic complete (SC), which consists only of the 22 amino acids for yeast together with sources of carbon (dextrose) and nitrogen (peptone)~\cite{CSHL_manual}. Four different fields of view were examined with approximately ten cells per field of view. The cells were imaged for 30 minutes at 5, 10, and 20 second exposure intervals.  In this minimal 
medium, a proportionate (expected) number of cells displayed anaphase spindle elongation, for all the exposure intervals studied. Furthermore, the labelled spindle poles were observed to maintain signal intensity longer when imaged in SC medium than in the lactate medium, further indicative of fewer oxygen free radicals being created from photon absorption.

Following these initial  studies, the spindle dynamics of cells incubated and imaged in SC medium was investigated in detail, using the protocol for quantitative image acquisition we derived from our studies.  
To minimize phototoxic effects, the acquisition interval was selected to be 10 seconds, and the total acquisition length to be 20 minutes. 
\subsubsection{Spindle dynamics in wild-type cells}\label{sec:results:spbsurftracking}   
A population of unsynchronized wild-type cells expressing the spc42-tdTomato reporter was incubated, washed, and imaged in SC medium. A confocal stack was collected with the 491 nm excitation line for each time point ($10\,\mathrm{s}$ sampling rate), with the exposure time set to $50\,\mathrm{ms}$. 
Within a given field-of-view, only cells for which two spindle poles could be detected within our resolution limits were analyzed. Signal-to-noise levels were typically between 7 and 11 dB. The spindle pole dynamics were analyzed in all cells in three-dimensions. In these data, those cells with two poles were segregated from the data.  For each time-point in these cells, the displacement between the two poles was computed, to construct the time evolution of spindle length. For cells in metaphase and early to mid-anaphase, the separation between SPB's corresponds to the spindle length. An additional level of segregation based on spindle length can thus be performed to yield separate metaphase and anaphase subpopulations. 
For cells in the latest stage of anaphase, the elongated spindles begin to buckle and then disassemble as a mechanical entity,
so that in this latest phase of anaphase, the SPB separation does not reflect the true, curvilinear, intact spindle length. 

The measured spindle pole separation versus time obtained following analysis of the resulting images is plotted in Fig.~\ref{fig:results:SC_WTtraj}. For representative cells from the total imaged population, individual spindle-length trajectories across the entire imaging time are displayed.
\begin{figure}[ht]
  \begin{center}
 \includegraphics[width=0.48\textwidth]{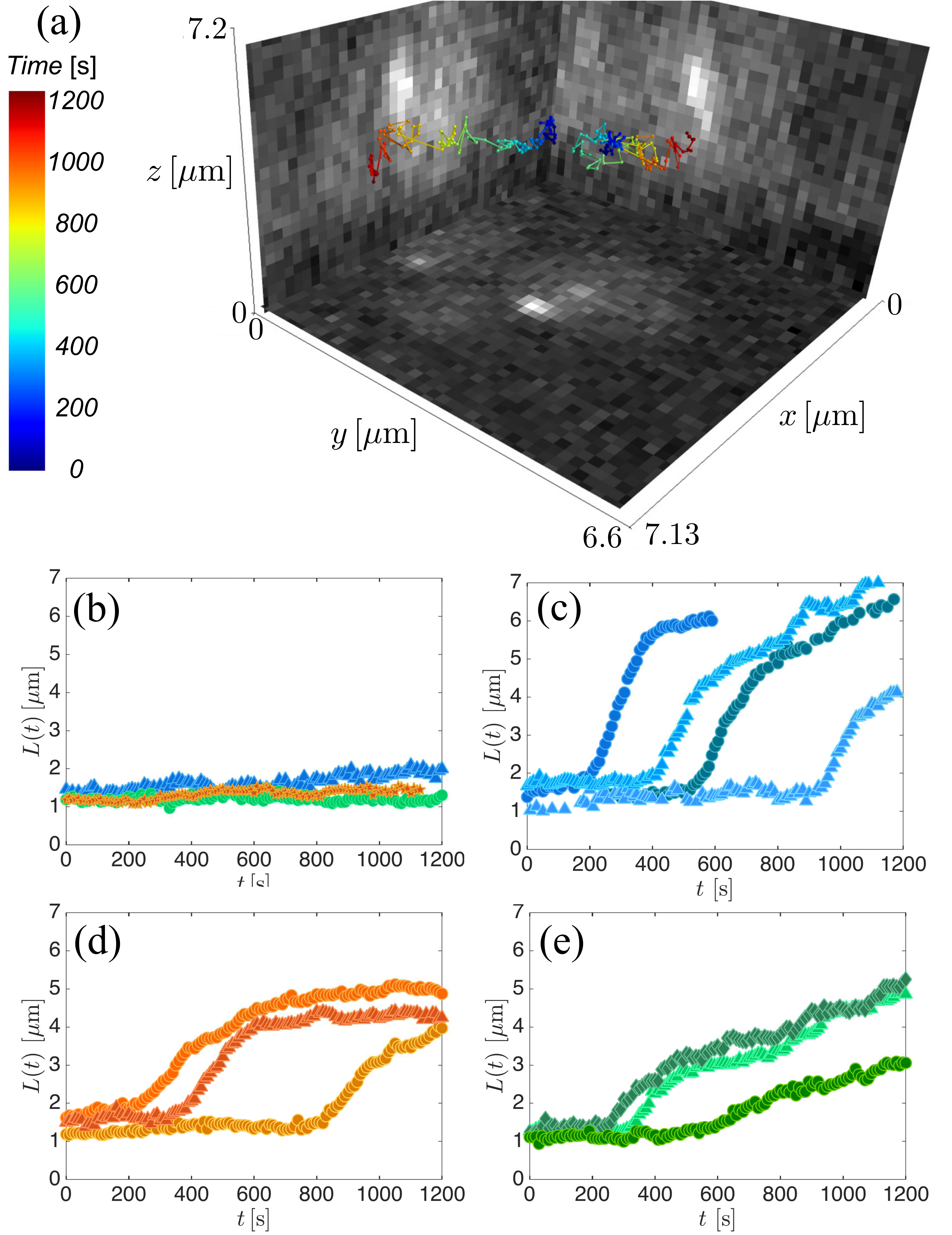}
    \caption{(a) The rotationally and translationally invariant spindle pole-spindle pole distance $L$ as a function of time, in a cell traversing anaphase. 
 (b)  Results are displayed for 3 metaphase cells showing representative bipolar spindle behaviors from a total population of 
  $\sim100$ analyzed cells each of wild type (blue triangles), kip1$\Delta$ (orange stars), and cin8$\Delta$ (green circles). (c-e) Representative anaphase behaviors of  (c), wild type cells; (d), kip1$\Delta$ cells; and (e), cin8$\Delta$ cells.}
  \label{fig:results:SC_WTtraj}
  \vspace{-0.8cm}
\end{center} 
\end{figure}
These data obtained from cells imaged in SC medium demonstrate 
the full range of stereotypical dynamical phases 
that are characteristic of normal progression through 
the cell cycle.
There are represented the approximately steady-state pole-pole separation at $1.5-2\,\mu\mathrm{m}$ characteristic of yeast metaphase (sometimes called pre-anaphase)(Fig.~\ref{fig:results:SC_WTtraj} (b)). 
There are the directed, apparently irreversible rapid spindle elongation, characteristic of yeast early anaphase~\cite{Straight_Science_97}, in which pole-pole separation increases rapidly 
% and smoothly 
from $1-2\,\mu\mathrm{m}$ to $5-7\,\mu\mathrm{m}$ over a duration of several minutes, followed by slower, less robust spindle elongation (Fig.~\ref{fig:results:SC_WTtraj} (c)). There are additionally cells with initially long spindles, in late anaphase, that exhibit 
uncorrelated pole-pole motion
characteristic of total spindle disassembly once anaphase. 

Using these tools and in SC medium, 
assignment of these cell cycle phases in an unsynchronized population may be done via automated segregation of individual traces based on their dynamical behavior. Since in budding yeast, chromosome condensation and a metaphase plate are not visible by microscopy, this  segregation provides a valuable intrinsic metric for cell cycle state.

In their performing of useful work in the cell, it is expected that the ensembles of kinesin-5 motors exhibit collective dynamics under nonequilibrium conditions. This should manifest itself in the observable $L(t)$. For cells in metaphase, the  SPB separation exhibited fluctuations superimposed upon a nearly constant, slowly increasing, function time. These fluctuations were well above the resolution limits of our feature finding of $10 - 13$ nm, as the signal to noise level over the imaging duration remained $\ge 8\,\mathrm{dB}$.

For cells in anaphase, the motor activity at the spindle mid-zone is increased~\cite{Khmelinskii-DevCell-2009,AvunieMasalaHoytGheber}. Due to their intrinsic directionality on the antiparallel MTs at the spindle mid zone, the extensile, coherent dipole character of the motors becomes manifest as directed, biased motion. At the scale of the whole spindle, this results in elongation. The collective dynamics of the motors gives rise coherent motion over longer timescales than the individual motors typically exhibit~\cite{Schmidt_Gerber_kip1_2013,Gerson-Gurwitz-Schmidt-Gheber}.
With the precision we obtain for the pole-pole distance in 3D, we can determine whether  fluctuations also appear superimposed on the directed, biased anaphase motion, as they do on the metaphase motion. 
%If so, the fluctuations could be extracted by subtracting off the fitted functional form that describes each cell.  
Upon observation of the motions with the detail enabled here, the anaphase motions were seen to be coherent, absent the tens of nanometer scale fluctuations we observed in metaphase. 
% This is not noise.

To investigate the specific contributions of mitotic motors to spindle dynamics, genetic perturbations of the mitotic motors were induced in budding yeast. Imaging and analysis of spindle dynamics using these tools was carried out for cells in which the gene for one or the other of the two kinesin-5 mitotic motors, $\rm{kip1}$ and $\rm{cin8}$, was deleted. Aside from these single-gene deletions, the genotypes were identical to that of the unperturbed cells studied above. Populations of cells in which one kinesin-5 motor was deleted were prepared with the $\rm{spc}42-\rm{tdTomato}$ fluorescent label. These $\rm{kip1} \Delta$ or $\rm{cin8}\Delta$ cell populations were cultured and imaged in the SC medium as described above. 
% poles in mutants
Data for the measured spindle pole separation versus time, obtained following analysis of the resulting images, for  representative cells from the total imaged populations, are plotted in Fig.~\ref{fig:results:SC_WTtraj}. 
The metaphase spindle length fluctuations displayed in Fig.~\ref{fig:results:SC_WTtraj} (b) do not differ appreciably between wild-type and mitotic motor mutant populations. 

Many of the cells are observed to pass through anaphase spindle elongation in both the kip1 and cin8 deletion strains. 
The nature of the anaphase motion in kip1-deleted cells, i.e. those carrying only the cin8 kinesin-5, is very similar to that in wild type cells, aside from slower spindle pole separation in late anaphase. 
By contrast, 
the cin8-deleted cells, i.e. those carrying only the kip1 kinesin-5, move more slowly through anaphase, consistent with what has been observed previously~\cite{Murray_JCellBio_98}. Additionally, this population also shows much greater variability in overall rate of SPB-SPB separation from cell to cell (data not shown). 

The nanometer-scale-resolution enabled by the tools we developed here permit more detailed examination. Upon further scrutiny,  
we see that the data for anaphase in $\rm{cin}8\Delta$ cells show some stop-start stuttering, where the extensile motion is apparently persistent for several tens of seconds, punctuated by periods of motion in the opposite direction. This oppositely-directed, contractile-like motion is not apparent in the anaphase motions of wild-type and $\rm{kip1} \Delta$  cells. It could arise from either brief, passive partial collapse of the spindle, or coherent (persistent) motion in the opposite direction; that is, active contractile motion after directional switching. Directional switching has  recently been observed for the kip1 motor {\em in vitro} under certain conditions~\cite{Schmidt_Gerber_kip1_2013}. 

We have recently used these context-rich tracking tools to examine the statistical mechanics of the fluctuations during metaphase for a large population of cells, in the context of models for the non-equilibrium activity of motors/microtubule (de)polymerization~\cite{Smith}.  
It will be interesting to examine the statistical mechanics of the coherent motions during anaphase, for populations of non-genetically perturbed cells and populations of cells with mitotic kinesin-5 molecular motors deleted, enabled by these methods, and to compare these observation to {\em in vitro} experiments once  {\em in vitro} systems are developed in which controllable {\em ensembles} of kinesin-5 motors can be studied~\cite{Spudich-minimuscle}.  

\subsubsection{Determining the cell cortex}\label{sec:results:spbsurftracking_mmutants}
\begin{figure}[ht]
  \begin{center}
  \includegraphics[width=0.48\textwidth]{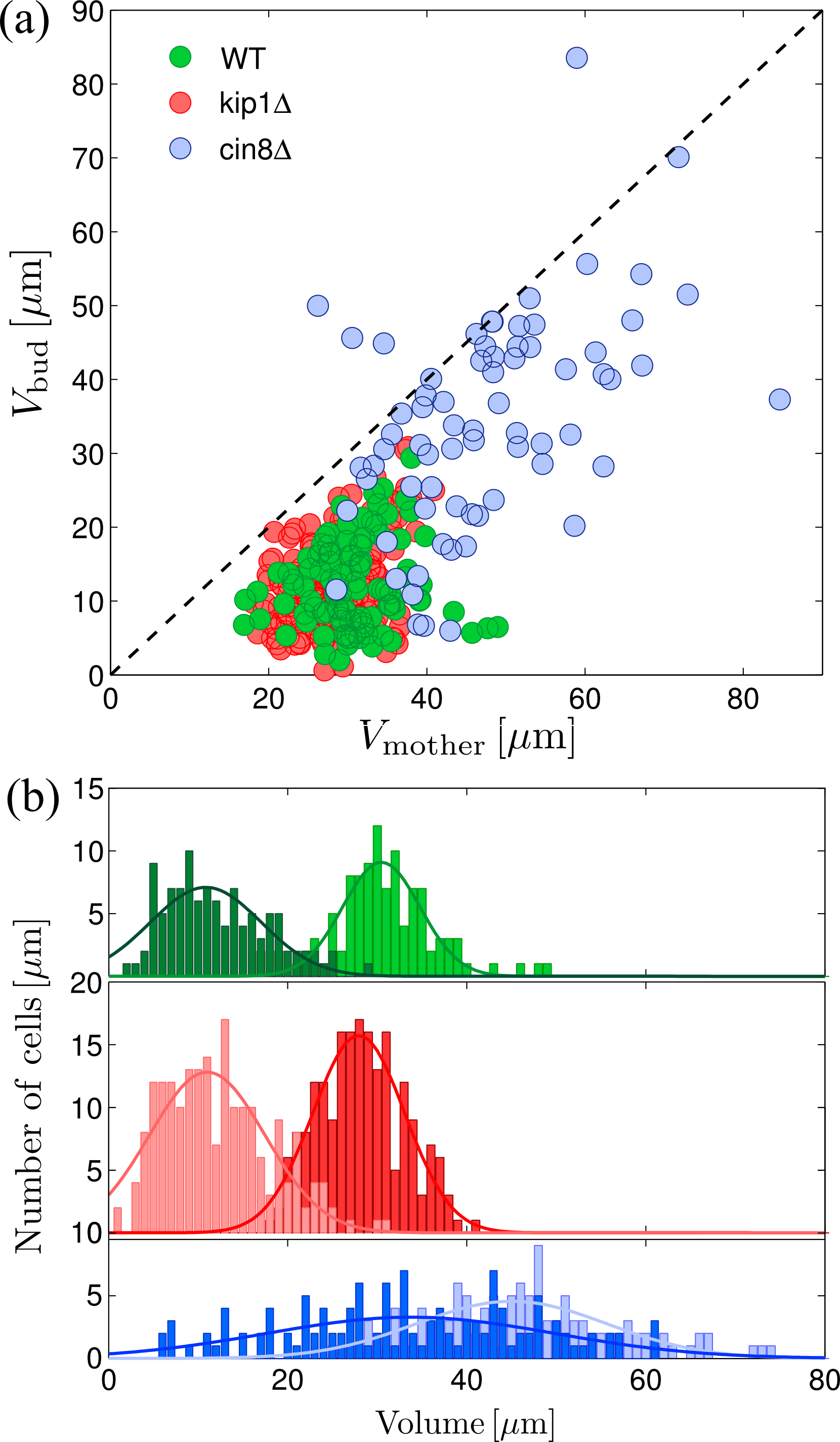}
    \caption{Top: Mother and bud cavity volumes determined from ellipsoid fits. Each bud volume is plotted against the volume of its mother cavity for wild-type cells (green) and for populations of cells with mitotic kinesin deletions $\rm{kip1}\Delta\,$ (red) or $\rm{cin8}\Delta\,$ (blue). The line at $V_{\mathrm{b}}=V_{\mathrm{m}}$ is drawn for reference. Bottom: Probability distributions of bud and mother cavity volumes for populations of pre-anaphase wild-type cells (dark green, bud; light green, mother) and for populations of cells with mitotic kinesin deletions ($\rm{kip1}\Delta\,$: bud, light red; mother, dark red; $\rm{cin8}\Delta\,$: bud, dark blue; mother, light blue).}
  \label{fig:results:mutant_volumes-dists}
\end{center} 
\end{figure}

Mother and bud compartment volumes of dividing cells were calculated using the method described above applied to the thinned data. In Fig.~\ref{fig:results:mutant_volumes-dists}, the bud cavity volume is plotted versus the mother cavity volume. Each data point corresponds to a separate wild-type cell. As expected, the bud cavity was always smaller than the mother cavity \cite{Hartwell_JCellBio_77}. In Fig.~\ref{fig:results:mutant_volumes-dists}, the mother and bud volume distributions over the cell population are plotted. Each of the mother and bud volume distributions were fit to a Gaussian function.
There is a degree of variability in the volume of the mother cell and in the volume of the daughter cell within a cell population, as was observed in early quantification using area as a proxy for volume~\cite{Hartwell_JCellBio_77}. Despite the variability, the distributions are bounded from above and below, due to known mechanisms of the cell cycle. 
The distribution of mother cells is bounded on the lower side by a minimum  cell size threshold for replicative division; and on the upper side by the fact that the time from one division to the next is compressed in cells that have already divided, and by the fact that  during mitosis, growth of the mother cell is minimal as most of the increase in mass goes into bud growth~\cite{Hartwell_JCellBio_77}. 

% volumes in mutants
Cell volumes quantitated for the motor deletion strains are displayed in Fig.~\ref{fig:results:mutant_volumes-dists}, superimposed on the results for wild-type cells.
The data show that $\rm{kip1}\Delta\,$  cells have mother and bud volume distributions practically unchanged from those of wild-type cells.  By contrast, 
% In the distribution of cavity volumes in Fig.~\ref{fig:results:mutant_volumes-dists}, 
the volumes in the $\rm{cin8}\Delta\,$ population are markedly larger, for both the mother and bud cavities, than in wild-type and $\rm{kip1}\Delta\,$ populations. Moreover, the distribution of $\rm{cin8}\Delta$ volumes across the population is  much broader for both mother and bud, and the mother and bud distributions are no longer distinguishable. 
% The volume distinction between the two cavities is blurred. 
The mean and standard deviations of mother and bud cavity volumes obtained from Gaussian fits to all the distributions are collected in Table \ref{tab:volumes}. 
\begin{table}[htbp]
  \begin{center}
    \caption{Mean $\mu$ and standard deviation $\sigma$ of the distribution of mother and bud volumes over populations of $N$ cells of wild-type cells or with one or the other of the mitotic kinesin-5 motor proteins deleted.}
\label{tab:volumes}
    \begin{tabular}{c|ccccc}
      \hline
      {volumes in $\mu\rm{m}^3$} & $\mu_{\mathrm{mother}}$ & $\sigma_{\mathrm{mother}}$  &  $\mu_{\mathrm{bud}}$  &  $\sigma_{\mathrm{bud}}$ & $N$ \\
      \hline
      wild-type & 30.46  & 4.33 & 10.91 & 6.05 & 160 \\
      $\rm{kip1}\Delta$ &  27.94 & 5.14 & 11.05 &  6.30 & 199 \\
      $\rm{cin8}\Delta$ & 45.05 & 10.97   & 32.96 &  14.92 & 124 \\
      \hline
    \end{tabular}
    \end{center}
\end{table}

The differences observed between wild-type and motor-deleted populations 
on the large mother and bud cavity volumes 
suggests a potential relationship between the control mechanism for dynamics of chromosome segregation, in which the kinesin-5 motors are involved, and that for cell size. Most striking is the effect of $\rm{cin8}$ deletions on cell volume. 
It is possible that 
% timing alone of two independent processes is the relationship: 
the large volumes observed in the $\rm{cin8}\Delta\,$ population were a result of a delay in the cell cycle for this population during a stage of cell growth: If chromosome segregation in the mitotic spindle is delayed, the cell may continue to produce and partition membrane and mass to the bud, resulting in a bud much larger than normal; eventually, the distribution stabilizes in homeostatic equilibrium after multiple generations. 
Nevertheless, there is a remarkable correlation, synchrony. Given the synchrony already observed between sister chromatid separation and onset of rapid spindle pole separation, specifically at the molecular level by increased coherent kinesin-5 motion in the spindle midzone.  
This synchrony is now known to be due sharing of the {\em same} signaling elements across the two processes~\cite{HoltMorgan-sharpanaphaseswitch-2008}.
It would not be surprising if cell wall growth were also shown to be in synchrony through some shared feedback mechanism with mitotic motors on the spindle. 

\begin{figure}[ht]
  \begin{center}
	\includegraphics[width=0.53\textwidth]{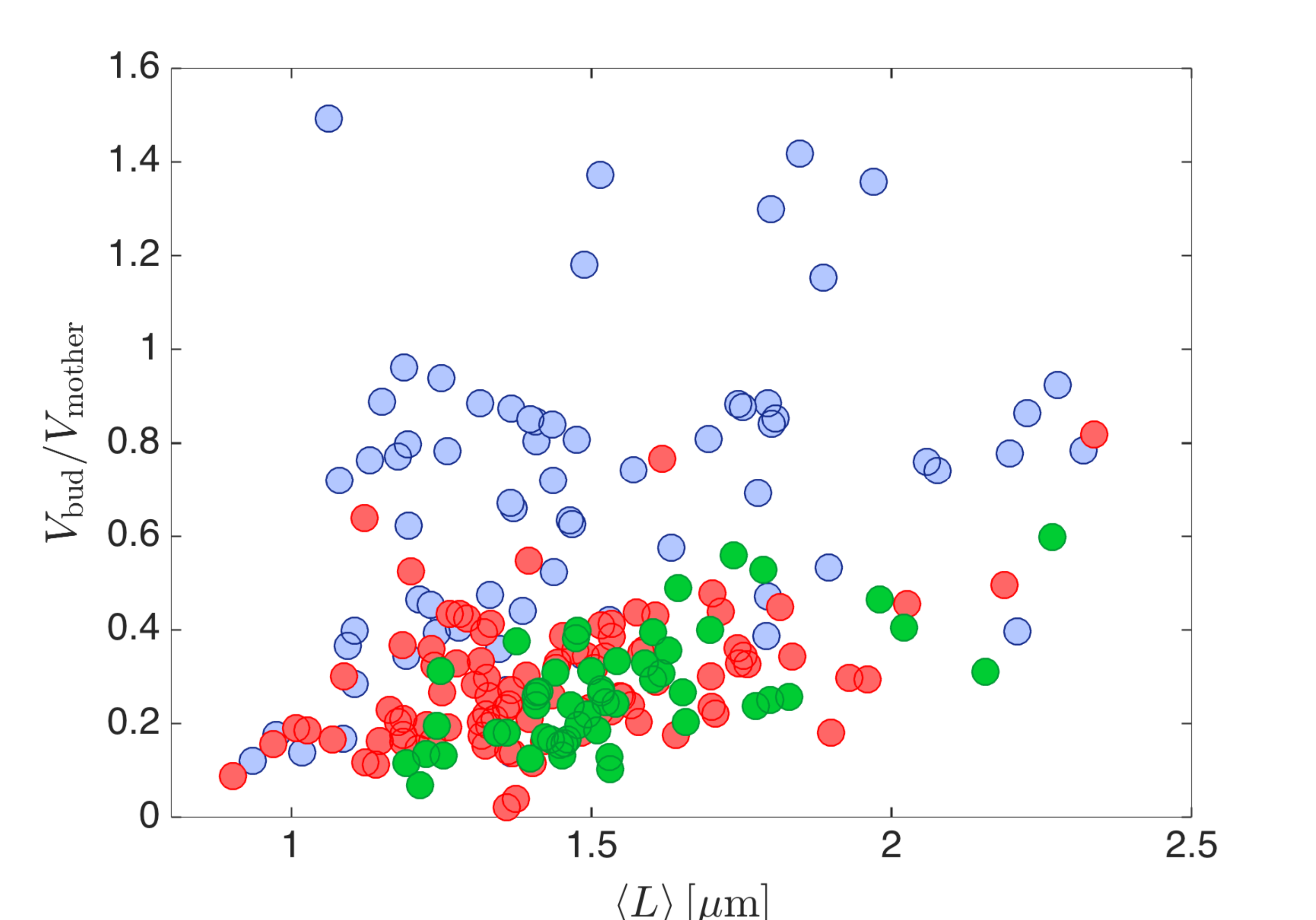}
    \caption{Ratio of bud-to-mother cavity volume versus mean spindle length for populations of pre-anaphase cells of wild-type  (green) and of type with mitotic kinesin $\rm{cin8}\Delta\,$ (red) or $\rm{cin8}\Delta\,$ (blue) deleted.}
  \label{fig:results:volRatio_vs_meanL_overlay}
\end{center} 
\end{figure}

% Combining metrics
Since cell growth is coordinated with the cell cycle~\cite{Hartwell_JCellBio_77}, the bud-to-mother volume ratio serves as a complementary metric for cell cycle progression that at a larger spatial scale than the spindle.
To explore this further, we demonstrated here use the tools we developed in tandem. 
 In Fig. \ref{fig:results:volRatio_vs_meanL_overlay}, the bud-to-mother volume ratio is plotted against the mean SBP separation $\langle L\rangle$. For wild-type pre-anaphase cells, the bud-to-mother volume ratio is approximately proportional to $\langle L\rangle$, indicating a correlation between cell growth and mitotic progression. 
The data suggest that spindle length and bud-to-mother volume ratio are coupled: there is a constant ratio between nuclear size and cell size. 
There may be global mechanical signals that tell a cell or organelle about its overall shape.

 The coupling between mitotic progression and cell growth appears to be a fundamental property of wild-type pre-anaphase cell populations~\cite{JorgFutc07}. Although a link between nuclear size and cytoplasmic volume has been suspected for many years, recent studies in yeast have introduced a tractable genetic system in which this question could be answered. 
 
 Fig.~\ref{fig:results:volRatio_vs_meanL_overlay} also plots the bud-to-mother volume ratio versus $\langle L\rangle$ for the motor-deleted populations. In the $\rm{kip1}\Delta\,$ cell population, the  bud-to-mother volume ratio versus $\langle L\rangle$ follows the same trend as for wild-type cells. 
For the pre-anaphase $\rm{cin8}\Delta\,$ population, the bud-to-mother volume ratio is dramatically less coupled to $\langle L\rangle$ as compared to wild-type and $\rm{kip1}\Delta\,$ cells. In some $\rm{cin8}\Delta\,$ cells, the bud has grown larger than the mother, exhibited by the values of the bud-to-mother volume ratio greater than one. 

Removal of the cin8 motor may perturb components of the control system, altering the coupling between chromosome segregation and cell size control. This can help to explain how the bud cavity may  become even larger than the mother cavity. 
It suggests that if there are internal mechanical signals that tell 
a cell about its overall state at a larger length scale, these signals are perturbed when the composition of certain force-generating elements of the mitotic spindle are perturbed.

Although cell size control in yeast has been explained for decades in terms of a minimum cell size before entry into mitosis~\cite{Hartwell_JCellBio_77}, many questions remain. For example, it is unclear how yeast--which lack lamins and lamin-associated proteins--adjust nuclear volume in response to changes in cytoplasmic volume.  In any organism, the mechanism by which the upper limit to nuclear growth is established is unknown. 

The results presented here serve as a proof of principle of the methods developed to track spindle pole bodies  with high accuracy in full 3-D space, and faithfully reconstruct budding yeast cell surfaces and volumes. 

\section{Conclusion}
%
% such as XXX, using a confocal microscope is that conventional instruments are only capable of XXX very small areas at unknown macroscopic location in the sample. In this work, a XXX high resolution tracking of nanoscale spots in living cells at identifiable locations on a mesoscopic scale (integrated surface and point finding) using XXX is reported. 
%
%In future, it would be interesting to forego the pruning step described here and apply directly to the unfiltered data a refinement of the algorithm for noisy data~\cite{robustCocone}.

%The above results validate the use of XXX for intracellular nanometer-scale resolution imaging, and the method can, in principle, be used in the investigation of endogenous intracellular molecules, provided that they can be targeted by function- alized QDs inserted in the cytoplasm. Our measurements therefore represent a first step toward experiments in which QDs, with distinct emission colors and bound to specific biomolecules (either in the membrane, the cytoplasm, or the nucleus), are tracked simultaneously and in real time. Depending on cell lines and molecular targets, this will possibly require nanoparticles with smaller hydrodynamic radii. We thus anticipate that fine adjustment of the QD colloidal properties through their surface coating will be critical for the investigation of intracellular processes at the single-particle level.
%

We have presented of a novel application of high-resolution position finding in conjunction with computational geometry techniques to reconstruct surfaces of cells as small as a few micrometers, to locate and track dynamic entities at their address in a living cell, and demonstrated the use of this method in capturing subcellular dynamics. 
%
%The development of a novel application of surface reconstruction and decomposition presented here, in conjunction with the high-resolution position finding, permits the measurement of nanometer scale motions of microscopic structures relative to each other or a reference point on the surface, removing sources of noise. 
%Here, we use these methods to demonstrate nanometer-scale measurements of directed motions or fluctuations of the eukaryotic mitotic spindle, using relative motion to unambiguously segregate cells in different stages of cell division; and to demonstrate measurements of cell volume of individual small biological cells.
%For mitosis to be tractable, the key will 
%be understanding how 
%mechanical communication 
%between the molecules regulates 
%and coordinates the spindle, which will require accurate measurements of motions and forces of the molecules. To measure motions and forces in this process, methods are needed to track the spindle dynamics of single cells in three dimensions. 
% The site of organelles or of the spindle poles within the bounding volume of the cell is usually not precisely known. 
Besides imaging features at high resolution in context-rich fashion, our technique also makes possible measurement of the volume and mapping of the surface of small individual cells or subnuclear regions for which the signal from the bounding surface may suffer from local reporter heterogeneity or high underlying surface curvature. 
The ability to calculate volumes and shapes in situations of low sampling of surface points is likely to become more important as we develop more sophisticated models of how the interior of the cell and the interior of the nucleus are organized, and how they deform and transduce forces.

\begin{acknowledgments}
The authors thank Professor Tamal K. Dey, Department of Computer Science and Engineering, Ohio State University for providing us surface mesh generation software, and Dr. Jeffrey N. Strathern, NIH/NCI, for providing the yeast strain carrying the spc42-tdTomato fusion protein.
\end{acknowledgments}

%%\bibliographystyle{unsrt}
% \bibliography{./Refs/references}

%%%%%%%%%% Merge with supplemental materials %%%%%%%%%%
\widetext
\clearpage
\begin{center}
\textbf{\large Supplemental Materials: Automated three-dimensional single cell phenotyping of spindle dynamics, cell shape, and volume}
\end{center}
%%%%%%%%%% Merge with supplemental materials %%%%%%%%%%
%%%%%%%%%% Prefix a "S" to all equations, figures, tables and reset the counter %%%%%%%%%%
\setcounter{equation}{0}
\setcounter{figure}{0}
\setcounter{table}{0}
\setcounter{page}{1}
\setcounter{section}{0}
\makeatletter
\renewcommand{\theequation}{S\arabic{equation}}
\renewcommand{\thefigure}{S\arabic{figure}}
\renewcommand{\thesection}{S\arabic{section}}
\renewcommand{\bibnumfmt}[1]{[S#1]}
\renewcommand{\citenumfont}[1]{S#1}
%%%%%%%%%% Prefix a "S" to all equations, figures, tables and reset the counter %%%%%%%%%%

\section{Characterization of the imaging system}
\begin{figure}[ht]
  \begin{center}
   \includegraphics[width=0.95\textwidth]{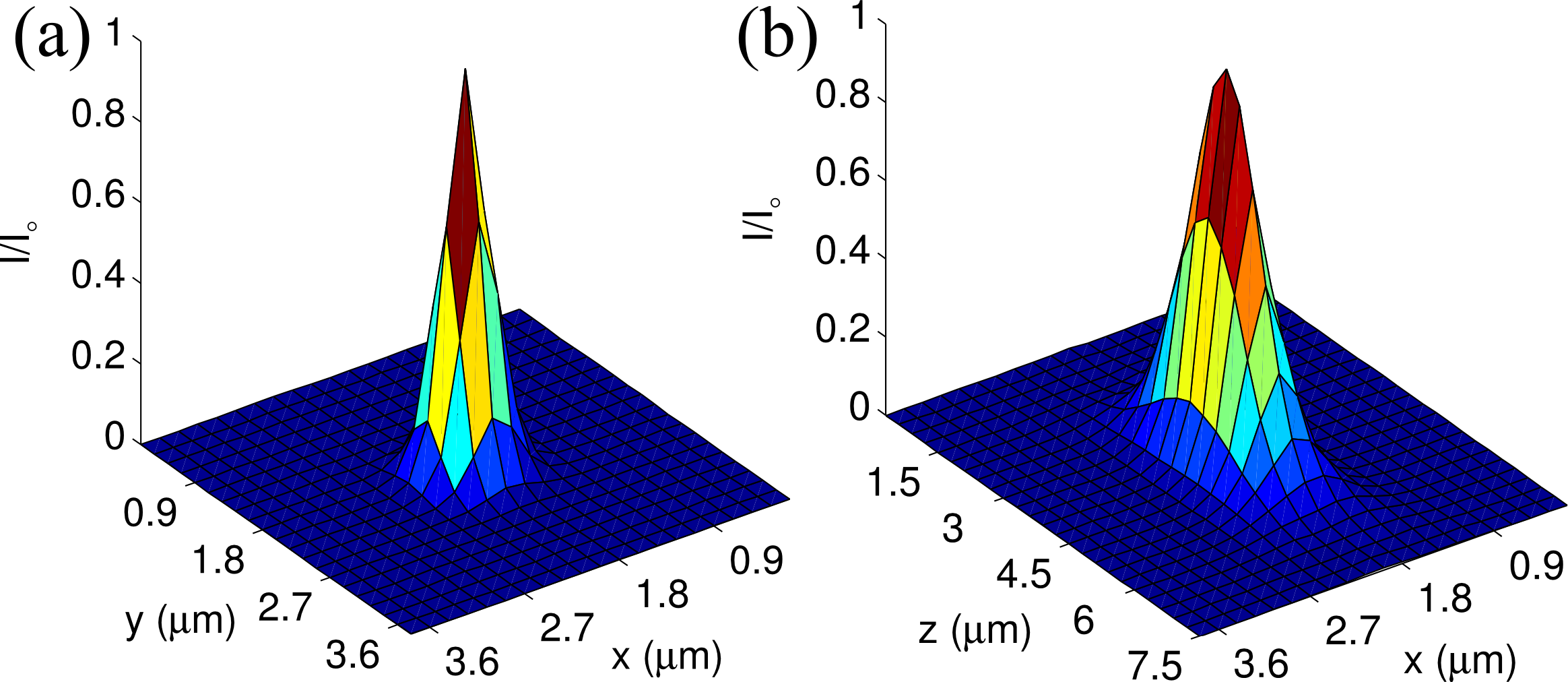}
   \caption{The measured three-dimensional PSF. (a) The centre $(x,y)$ plane, and (b) the centre $(x,z)$ plane, of the three-dimensional point spread function measured by acquiring, localizing, and averaging intensity distributions of 100 nm diameter fluorescently labelled beads. The PSF for the $\lambda_{ex}=491\,\mathrm{nm}$ channel shown here. The PSF for $\lambda_{ex}=532\,\mathrm{nm}$ channel  is  similar (see text).}
  \label{fig:psf}
\end{center}  
\end{figure}

To measure the true PSF of the microscope, sub-diffraction beads were imaged in 3D with a confocal point scanner (VT-Eye, VisiTech international) attached to an inverted microscope (Leica DM4000) with a $100\times$ oil-immersion objective (Leica Microsystems HCX PL APO, numerical aperture of 1.4) at wavelengths $491\,\mathrm{nm}$ and $532\,\mathrm{nm}$. Sub-resolution $0.100\,\mu\mathrm{m}$ diameter beads (TetraSpec, Molecular Probes, Eugene, OR) were imaged at $60\,\mathrm{ms}$ exposure time. Coverslips were coated with $0.1\,\mathrm{mg/ml}$ poly-D-lysine to adhere beads to the coverslip. The beads remained immobile throughout the experiment. The concentration of beads was sufficiently low that they were spaced far apart, and only 15 to 20 beads were visible in a given field-of-view ($ 90\,\mu\rm{m} \times 90\,\mu\rm{m}$). For five different fields-of-view, a set of confocal image stacks was acquired at a spacing of 0.2 \um between confocal planes. Typically, 40 stacks were acquired per set. To estimate the noise-free PSF, the intensity distribution of each bead in each of the image stacks was localized to sub-pixel resolution using the feature finding algorithm described below. 
All intensity distributions for a single bead in a set were then averaged with their centroids aligned. Since noise is not temporally or spatially coherent, this averaging dramatically reduces the noise signal. $(x,y)$ and $(x,z)$ profiles of the PSF obtained in this manner for the $491\,\rm{nm}$ excitation channel are shown in Fig. \ref{fig:psf}.

The measured point spread function was symmetric about the $x$, $y$ and $z$ axes, and showed no significant optical aberrations. 
The measured PSF's were fit to a three-dimensional Gaussian distribution $ A\exp[-((x-\mu_x)^2/(2\sigma_x^2) + (y-\mu_y)^2/(2\sigma_y^2) + (z-\mu_z)^2/(2\sigma_z^2) )]$, to obtain a model PSF for simulations. Reported values are the mean and standard deviation of the Gaussian fit parameters, over all five different fields-of-view. The standard deviations of the Gaussian PSF obtained for the GFP ($\lambda_{ex}=491\,\mathrm{nm}$) channel were $1.002 \pm 0.012$ pixels, $0.992 \pm 0.011$ pixels, and $2.522 \pm 0.020$ pixels ($174\,\mathrm{nm}$, $173\,\mathrm{nm}$, and $757\,\mathrm{nm}$)  in the $x$, $y$, and $z$ directions, respectively. For the tdTomato ($\lambda_{ex}=532\,\mathrm{nm}$) channel, the measured standard deviations of the Gaussian PSF were $1.105 \pm 0.007$ pixels, $1.104 \pm 0.018$ pixels, and $2.838 \pm 0.016$ pixels ($192\,\mathrm{nm}$, $192\,\mathrm{nm}$, and $851\,\mathrm{nm}$) in the $x$, $y$, and $z$ directions, respectively.

The readout noise cannot properly be determined by calculating a histogram for all pixel intensities across a single image because fixed-pattern noise, whose sources are pixel-to-pixel variations in dark current and non-uniformities in photoresponse~\cite{Pawley_ConfocalHanbook_Digitizing}, add additional spread to the intensity distribution. 

The background signal in the absence of fixed-pattern noise was obtained by using the intensity recorded for a single pixel over many measurements. To accurately reproduce the background present when imaging live cells, a sample of imaging medium (SC) was placed in the microscope during the measurements. The objective lens was focused to a point in the media just beyond the coverslip, and 3000 images of a single focal plane were acquired using the shortest possible exposure time ($\sim\!30$ ms). This was performed under three different conditions. In the dark field condition, all shutters were open but no illumination source was used. For GFP and tdTomato illumination conditions, the 491 nm and 532 nm excitation lasers were respectively turned on. Since no florescent reporter is present in the sample, the measured intensity distributions represent an estimate of the detector background signal including scattered light effects, as well as CCD readout and dark current noise. Because the relevant units for describing noise in the photon detector are electrons, the signal is converted from ADU's to electrons in the CCD by multiplying by the CCD gain factor of 5.8 electrons/ADU and dividing by the EM gain factor used in the acquisition. 

Detector noise from dark current is expected to vary as $\sigma_D=\sqrt{DT}$, where $D$ is the dark current and $T$ is the exposure time. For the camera used, $D = 0.01$ electrons/pixel/sec \cite{Hamamatsu_manual}, so $\sigma_D=0.02$ electrons, and the noise from dark current is expected to be negligible. Readout noise should be Gaussian distributed with a variance of $\sim\!1$ electron \cite{Hamamatsu_manual}. Histograms were calculated for a single pixel at different regions over the field-of-view for each illumination condition.
%, and these are shown in Figure \ref{fig:img_sim:bckgrnd_hist}. 
For the dark field, 491 nm illumination, and 532 nm illumination, no appreciable change was observed between the histograms from different regions across the field of view. The intensity distribution for each of these conditions follows a Gaussian profile, but with a small exponential tail. This exponential tail may be explained by multiplicative noise in the gain-register of the EM-CCD, which will produce an exponential distribution of pulse heights with many small pulses and few large ones \cite{Pawley_ConfocalHanbook_Digitizing}. 

\section{Simulation of spindle poles}
\begin{figure}[ht]
  \begin{center}
   \includegraphics[width=0.90\textwidth]{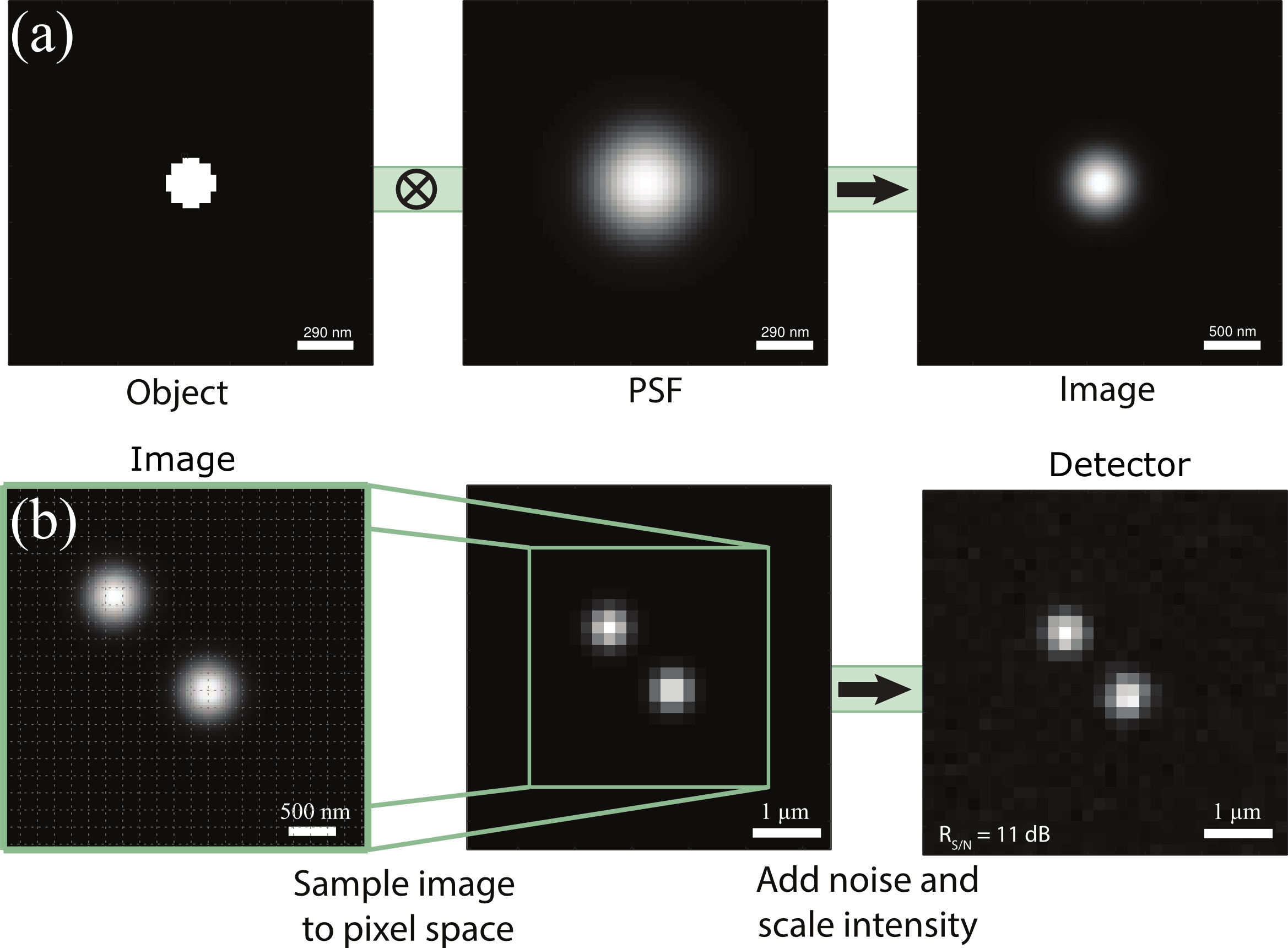}
   \caption{Our performance test data generation procedure at a glance: Image formation for a diffraction-limited spot, and simulation of pixelation and noise. 
   (a) Convolution of the object and PSF by fast Fourier transform and multiplication in Fourier space. The object and PSF arrays were zero-padded prior to Fourier transfomation to avoid circular convolution. (b) Image of  two diffraction-limited spots is sampled from the continuous image space to discrete pixel space on the CCD grid with added representative noise.}
  \label{fig:img_sim}
\end{center}  
\end{figure}

The image of a diffraction-limited spot was constructed by first generating a sphere of $150\,\mathrm{nm}$ diameter in a high resolution object space. The object space containing the sphere is a three-dimensional grid of voxels with dimensions of 29 nm in $x$ and $y$, and 50 nm in $z$. The choice of object space pixel size is a compromise between approximation of a continuous space, which requires small pixel sizes, and available computer memory, which limits the size of arrays that can be convolved in a practical time frame. A Gaussian approximation to the microscope PSF was also generated in the object space using the parameters found earlier. The Gaussian function used to generate the PSF was not normalized and the amplitude was set to 1 unit. Convolution of the object and PSF was performed by two-dimensional fast Fourier transform of each of the arrays, multiplication in the frequency domain, and finally, inverse Fourier transformation (Fig. \ref{fig:img_sim} (c)). To avoid circular convolution, the image and PSF arrays were appropriately zero padded prior to Fourier transformation. 

Convolution of the object with the PSF results in the intensity distribution of a diffraction-limited spot in the continuous image space, the magnitude of which is interpreted in terms of a number of photons. To generate a realistic image that is comparable to experimentally acquired images, it is necessary to sample the continuous image space intensity distribution into a discrete distribution collected by the CCD detector. To do this, the image space intensity distribution is resampled to a pixel grid, with each pixel representing an individual CCD element (Fig. \ref{fig:img_sim} (d)). To simulate the spindle poles, two diffraction-limited spots are produced in the image. The CCD grid has voxel dimensions of 174 nm in $x$ and $y$, and 300 nm in $z$, to mimic the experimental conditions. To construct the CCD voxel distribution, the continuous image space distribution was sampled at a number of focal planes in the $z$-stack. Finite depth-of-field was explicitly accounted for at each $z$-slice by weighting the image space matrix by a Gaussian function with an amplitude of 1 and a standard deviation of $2.8$ pixels, centered on the particular $z$-slice. The intensity distribution in the focal plane was then constructed by integrating the depth-of-field-weighted image space matrix along $z$. Each CCD pixel in the focal plane was formed by integrating the focal plane over square regions corresponding to the CCD pixels. The resulting matrix represents the distribution of photons impinging upon each of the CCD elements for each focal plane collected. This matrix is adjusted by a scaling factor to control the signal-to-noise. Sub-pixel displacements of the object were produced by shifting the image space grid by one or more of its elements, with respect to the overlying CCD grid.

Detector noise and background offset were added to each image stack at intensity levels corresponding to experimental values. To represent the photon detection process accurately, Poisson noise was simulated for each pixel by random sampling from a Poisson distribution with a mean of $N$, where $N$ represents the number of photons impinging on each pixel. The intensity units, which represent numbers of photons, were converted to grey scale units by multiplying by the quantum efficiency and by the gain factors of the CCD, and dividing by the photo-electron/ADU conversion factor provided in the camera manual \cite{Hamamatsu_manual}. These values could then be directly converted to integers, that represent grey scale levels in the camera's dynamic range (16 bit). The signal-to-noise ratio was calculated as the mean pixel value above background of the diffraction-limited spot divided by the standard deviation of the background pixel levels. The mean signal level $\mu_{signal}$, mean background level $\mu_{bkgrd}$, and variance $\sigma_{bkgrnd}^2$ were calculated in terms of numbers of photons, and the signal-to-noise ratio is expressed as a decibel quantity, 
\begin{equation}
R_{S/N}=10\log_{10}{\left(\frac{\mu_{signal}-\mu_{bkgrd}}{\sigma_{bkgrd}}\right)}.
\end{equation}

\section{Influence of filtering on localization of closely-spaced features}
\begin{figure}[ht]
  \begin{center}
    \includegraphics[width=0.8\textwidth]{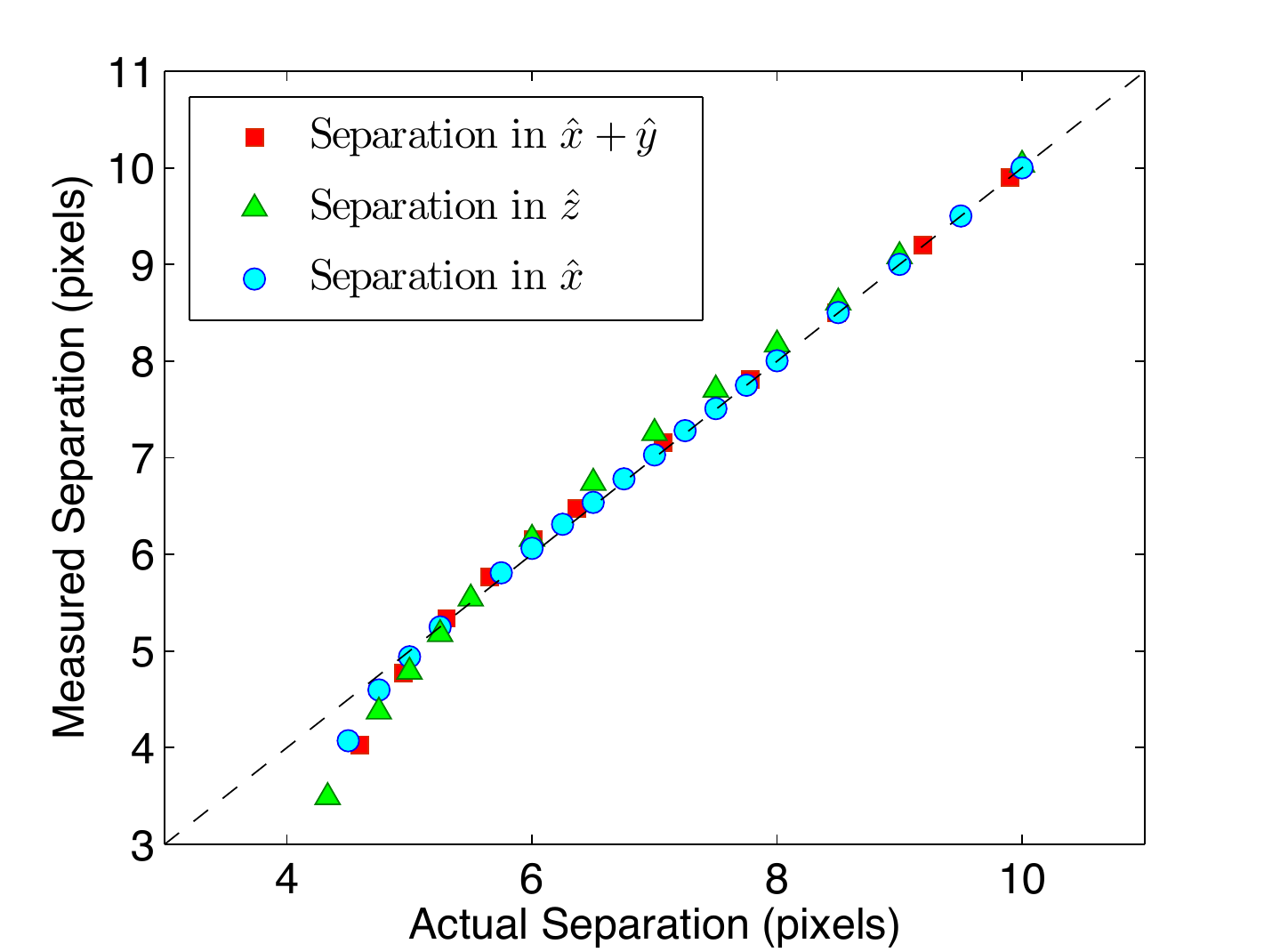}
   \caption{Point-to-point separation measured by the localization algorithm plotted against the true separation for a number of different displacement vector orientations. The dashed line indicates where measured separation equals the actual separation, for reference. Each data point is the average of 500 iterations. $R_{S/N}$ is $10.5$ dB.}
  \label{fig:img_sim:Msep_vs_Asep}
  % {Measured separation versus actual separation}
\end{center} 
\end{figure}

The localization error is not a monotonic function of the point-to-point separation. As the displacement decreases below $2r_{crit}$, there is a large increase in the localization error. At some separation below $2r_{crit}$, the error decreases, and then upon further reduction in separation rapidly diverges as the two features become difficult to distinguish. In order to examine this effect more closely, the measured point-to-point separation was compared to the true point-to-point separation. 
in Figure \ref{fig:img_sim:Msep_vs_Asep}.
%\begin{center}
%  \includegraphics[width=0.8\textwidth]{./figures/img_simulation/LFF_vs_LReal}
%\end{center}

We determined that the initial increase in error, at $\sim\!7$-9 pixels separation, is due to a systematic over-estimation of the point-to-point separation. The overlap between the intensity distributions forces the least-squares fitting routine to localize the centroid of each intensity distribution away from the overlap region, resulting in overestimation of the point-to-point separation. As overlap increases, the centroids of each feature are forced back towards the overlap region, initially decreasing the estimated separation, until significant overlap has occurred, at $\!\sim5$ pixels, and the two spots are difficult to distinguish. 

\begin{figure}[ht]
\begin{center} 
   \subfigure{	
      \includegraphics[width=0.48\textwidth]{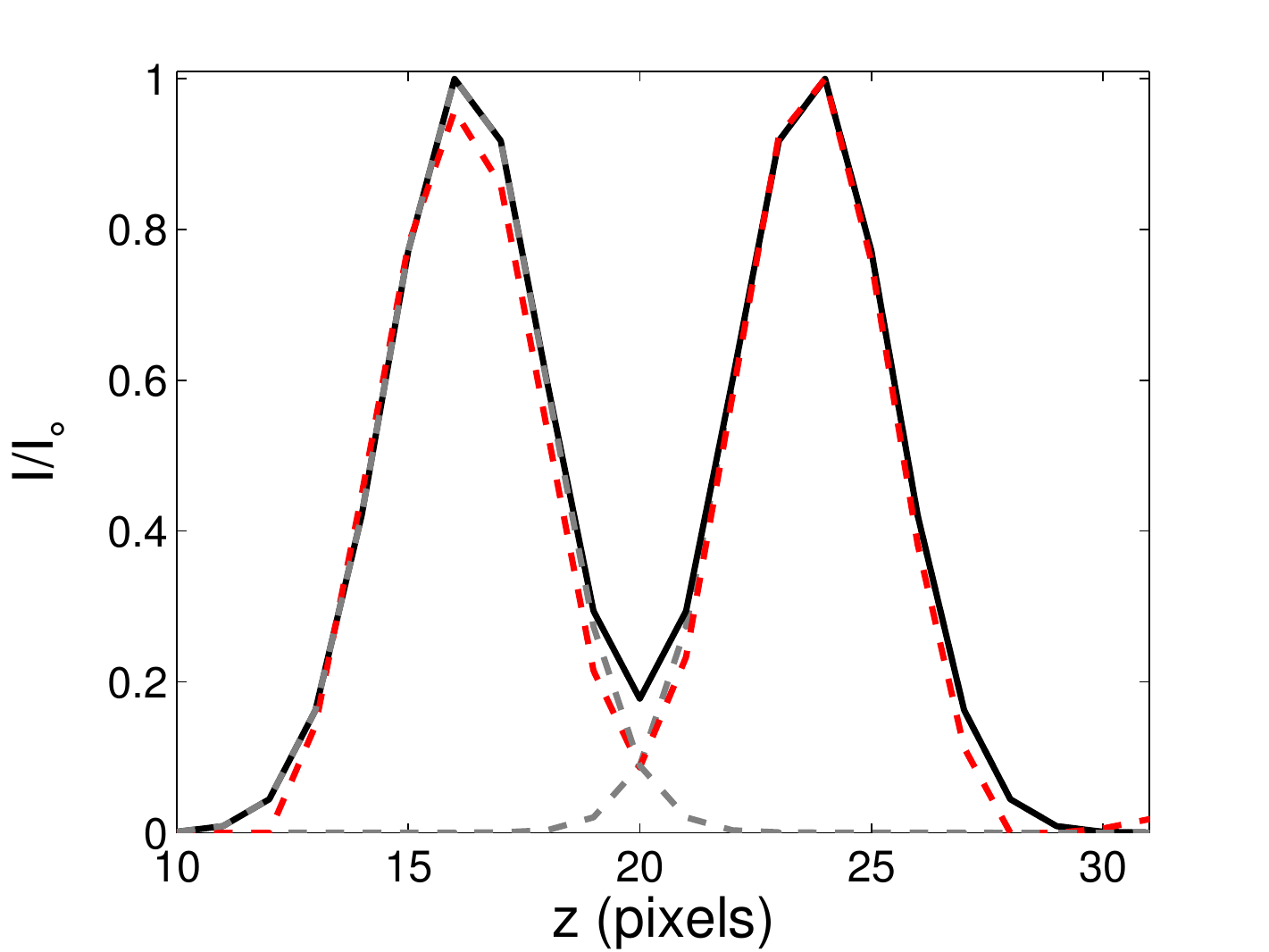}
    }
   \subfigure{
      \includegraphics[width=0.48\textwidth]{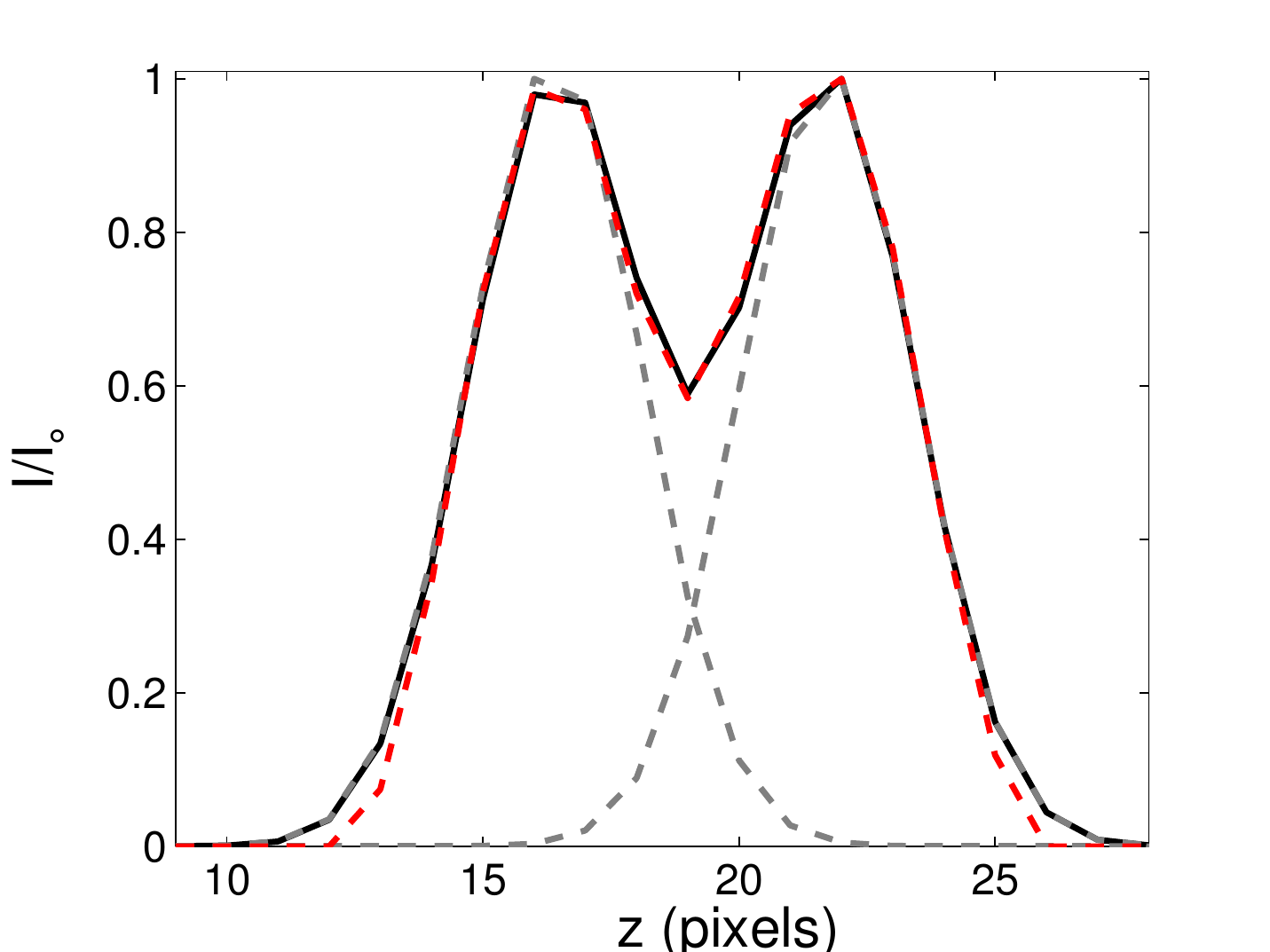}
     }
   \caption{Intensity distribution for closely spaced diffraction-limited spots. The intensity distributions along the axial, or $\hat{z}$ direction, for closely spaced diffraction-limited spots. In the left figure, the spots are centered at 16.25 and 23.75 pixels. In the right figure, the spots are centered at 16.42 and 21.75 pixels. Light grey dashed lines represent the intensity distributions from the individual points. Solid black lines represent the sum of the individual intensity distributions. Dashed red lines are the filtered intensity distribution. The effect of the band-pass filter can be seen in the region between the two spots and on the outside edges of the spots.}
  \label{fig:img_sim:2poinIntDist}
\end{center} 
\end{figure}

To investigate this behaviour of the localization error at close separations in greater detail, a simulation was performed wherein the features were localized in only one-dimension.  Three-dimensional images of point features were created as described above. The features were displaced from each other along the $\hat{z}$ direction by a known amount and the $z$ coordinate of each point was localized by fitting each intensity distribution independently to a one-dimensional Gaussian function. The centroids were displaced at fractional pixel values along the $z$ -axis and the integer pixel locations of the centroids were used as initial parameter estimates for the Gaussian fits. 
The $\hat{z}$-direction was chosen because localization of the $z$-coordinate has the largest error for closely spaced points. 
To localize features with limited spatial filtering,  i.e. the extreme values of \ $f_c \rightarrow \infty$ and $f_c =1$, the noise had to be reduced to unrealistically small values, mimicking unattainably perfect imaging conditions. For suitable comparison, 
the simulations in Figure \ref{fig:img_sim:2poinIntDist} were performed with $R_{S/N} \rightarrow \infty$. 

When filtering is used, it is a source of small overestimation of the separations. 
Filtering removes high spatial frequency components from the image, predominantly noise; however, the operation also smoothes components of the real intensity distribution that vary rapidly in space. When the distance between points is smaller than $2r_{crit}\sim 2 \sigma_{PSF}$, the filtering operation modifies the summed intensity distribution of the two features such that the distance between the two is overestimated by fitting each feature independently to a Gaussian function. 

  The origin of the behavior of the localization error for features that are closely spaced is verified upon examining the effect of the band-pass filter. 
At some separation close to $r_{crit}$, the effect of filtering in the overlap region becomes negligible compared to the effect of the overlap itself. 
At this separation the measured separation agrees with the actual separation again (Fig.~\ref{fig:img_sim:Msep_vs_Asep}). At these and smaller separations, the measured separation begins to systematically diverge from the actual separation, due to overlap of the feature PSF's.
%The one-dimensional simulations verify the behavior of the localization error found in the three-dimensional simulations. 
Plots of the intensity distributions before and after filtering for two different spacings are shown in Figure \ref{fig:img_sim:2poinIntDist}.

\section{Propagation of localization error}
\begin{figure}[ht]
  \begin{center}
     \includegraphics[width=0.6\textwidth]{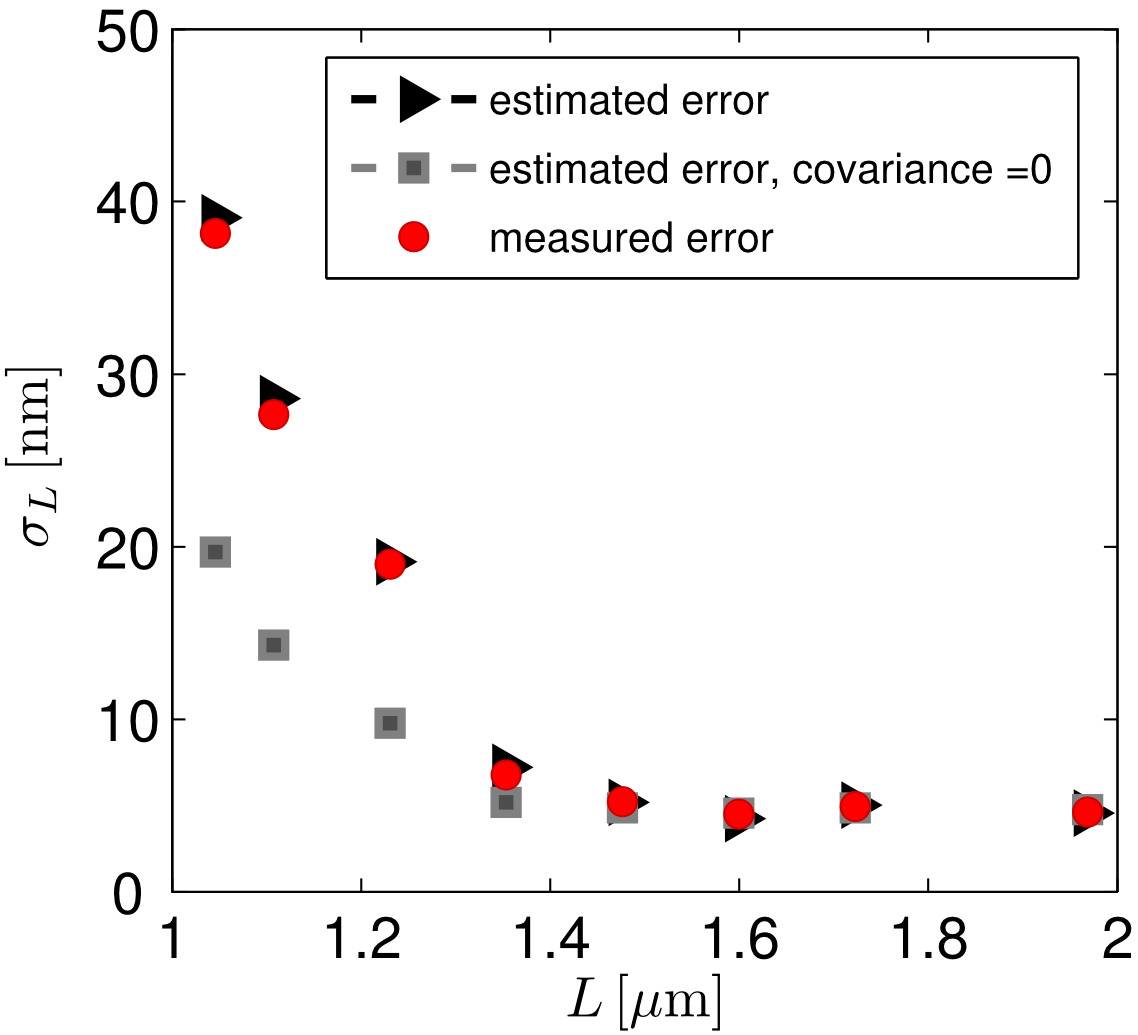}
   \caption{Error in point-to-point separation determined directly from simulation, compared with the error calculated by propagating through the errors in coordinate localization. Light grey dashed line is the propagated error neglecting covariance terms. Black dashed line is the error propagated with equation \ref{eq:errorProp}. $R_{S/N}$ for this simulation is 10.5 dB.}
  \label{fig:img_sim:ErrorPropCompare}
  % {Comparison of measured error with propagated error for point-to-point separation}
\end{center} 
\end{figure}

Propagation of localization error in the separation of two points was studied in order to determine its role, if any, in the observable $L(t)$, the spindle pole-pole separation. The error for any arbitrary quantity $\epsilon(x_i^a,x_i^b)$ that has been calculated from the set of coordinates for both features $\{x_i^a,x_i^b \}$ can be estimated by propagating the errors associated with each of the coordinates. For notational simplicity, the coordinates representing the positions of each of the features are written as the set $\{x_i\}$, where $i=\{1,2,...,6\}$. The errors for all the coordinates are written generally as $\{\sigma_{x_i},\sigma_{x_{ij}}\}$, and the covariance terms $\sigma_{x_{ij}}$ are given by 
%\[
$ \sigma_{x_{ij}}^2 = 1/N\sum\limits_{i,j}\left(x_i - \mu_{x_i}\right) \left(x_j - \mu_{x_j}\right)$.
% \]
The set of values $\{\mu_{x_i}\}$ is the set of most probable estimates of the true coordinates, which is exactly known for the simulations. Frequently when uncertainties are estimated, the covariant terms are assumed negligible. However, for parameters obtained from fitting a curve to data, the covariant terms can contribute significantly to uncertainties \cite{Bevington}. The appropriate relation for estimation of uncertainty in $\epsilon(x_i)$ is:
\begin{equation}
\label{eq:errorProp}
\sigma_\epsilon^2  \simeq  \sum\limits_{i}{\sigma_{x_i} \left(\pd{\epsilon}{x_i}\right)^2} + \sum\limits_{i\neq j}\sigma_{x_{ij}} \left(\pd{\epsilon}{x_{ij}}\right)^2.
\end{equation}
The error in point-to-point distance can be determined directly during a simulation by comparing the real separation with the measured separation. This error can also be determined from the uncertainties of each coordinate via equation \ref{eq:errorProp}, since $ L^2=\sum\limits_{i=1}^{3} (x^a_i-x^b_j)^2 $. To test this, a simulation was performed with the displacement vector between the points oriented in an arbitrary direction in $\hat{x}$, $\hat{y}$, and $\hat{z}$, and the error in $L$ determined over a range of separations. 
The results of this simulation are displayed in Figure \ref{fig:img_sim:ErrorPropCompare}.

It can be seen that for large separations, 
in our case for $L\ge1.4\,\mu\mathrm{m}$, 
the covariant terms are negligible. As the two features approach one another to within less than $1.4\,\mu\mathrm{m}$, the covariant terms become significant and they must be included for an accurate estimate of the error. By simulating point-to-point separation trajectories for each of the directions in the set $\{\hat{x},\hat{z},\hat{x} + \hat{y} + \hat{z}\}$ the form of all $\{\sigma_{x_i},\sigma_{x_{ij}}\}$, can be obtained. Using this relation for $\sigma_\epsilon^2$, the errors for all quantities determined from feature finding can be estimated. 

\section{Chromatic shift}
Chromatic shift was measured and corrected for prior to combining results obtained using two different wavelengths. The chromatic shift in $x$, $y$, and $z$ between the GFP and red acquisition channels was determined by imaging approximately $200$ $2.5\,\mu\rm{m}$-diameter InSpeck beads (Invitrogen) immobilized on a poly-K-coated cover glass. The intensity distribution of each of the beads in a field-of-view was fit to a three-dimensional Gaussian function to find the centroid of every bead in each channel. GFP- and red-channel centroid positions were then compared for each of the immobilized beads to determine the chromatic shift between the excitation channels, averaged across all the beads. 
% 
%%Figure \ref{fig:results:chrom_shift} displays the measured chromatic shift for all beads, plotted as a histogram. 
The $z$ position in the 532 nm channel was systematically higher than it was in the 491 nm channel by $221\,\rm{nm}$ for the 63X objective lens. The in-plane chromatic shift between the 532 nm and 491 nm channels was $64\,\rm{nm}$ and $-31 \,\rm{nm}$, in $x$ and $y$ dimensions, respectively. 
The measured chromatic shift is corrected for in all succeeding analysis where data from the two excitation channels is combined.

\end{document}